\documentclass{aa}  

\pdfoutput=1

\usepackage{caption}
\usepackage{graphicx}
\usepackage{natbib}
\usepackage{lscape}
\usepackage{rotating}
\bibpunct{(}{)}{;}{a}{}{,} 
\usepackage{txfonts}
\usepackage{lscape}
\usepackage{breqn}
\usepackage{url}
\usepackage{longtable}
\usepackage{hyperref}
\begin{document} 

\title{The HADES RV Programme with HARPS-N@TNG\thanks{Based on observations made with the Italian Telescopio Nazionale Galileo (TNG), operated on the island of La Palma by the Fundaci\'on Galileo Galilei of the INAF (Istituto Nazionale di Astrofisica) at the Spanish Observatorio del Roque de los Muchachos of the Instituto de Astrof\'isica de Canarias (IAC).}}
\subtitle{II. Data treatment and simulations} 

\author{M. Perger\inst{1} 
	\and A. Garc\'{\i}a-Piquer\inst{1}
	\and I. Ribas\inst{1}
	\and J.C. Morales\inst{1}
	\and L. Affer\inst{2}
	\and G. Micela\inst{2}
	\and M. Damasso\inst{3}
	\and A. Suárez-Mascareño\inst{4,5}
	\and J. I. González-Hernández\inst{4,5} 
	\and R. Rebolo\inst{4,5,6}
	\and E. Herrero\inst{1}
	\and A. Rosich\inst{1}
	\and M. Lafarga\inst{1}	
	\and A. Bignamini\inst{7}	
	\and A. Sozzetti\inst{3}	
	\and R. Claudi\inst{8}	
	\and R. Cosentino\inst{9}
	\and E. Molinari\inst{9}	
	\and J. Maldonado\inst{2}	
	\and A. Maggio\inst{2}		
	\and A. F. Lanza\inst{10}	
	\and E. Poretti\inst{11}
	\and I. Pagano\inst{10}
	\and S. Desidera\inst{7} 
	\and R. Gratton\inst{8}
	\and G. Piotto\inst{8,12}
	\and A. S. Bonomo\inst{3}
	\and A. F. Martinez Fiorenzano\inst{5}		
	\and P. Giacobbe\inst{3}		
	\and L. Malavolta\inst{8,12}				
	\and V. Nascimbeni\inst{8,12}						
	\and M. Rainer\inst{11}	
	\and G. Scandariato\inst{10}
	}

	\offprints{M. Perger, \email{perger@ieec.cat}}

   \institute{\inst{1}Institut de Ci\`encies de l'Espai (CSIC-IEEC), Campus UAB, Carrer de Can Magrans s/n, 08193 Cerdanyola del Vall\`es, Spain\\
   \inst{2}INAF - Osservatorio Astronomico di Palermo, Piazza del Parlamento 1, 90134 Palermo, Italy\\ 
   \inst{3}INAF - Osservatorio Astrofisico di Torino, via Osservatorio 20, 10025 Pino Torinese, Italy\\         
   \inst{4}Institute de Astrofísica de Canarias (IAC), 38205 La Laguna, Tenerife, Spain\\                
   \inst{5}Universidad de La Laguna (ULL), Dpto. Astrofísica, 38206 La Laguna, Tenerife, Spain\\	
   \inst{6}Consejo Superior de Investigaciones Científicas (CSIC), 28006, Madrid, Spain\\				
   \inst{7}INAF - Osservatorio Astronomico di Trieste, via Tiepolo 11, 34143, Trieste, Italy\\	
   \inst{8}INAF - Osservatorio Astronomico di Padova, Vicolo dell’Osservatorio 5, 35122, Padova, Italy\\                
   \inst{9}INAF - Fundación Galileo Galilei, Rambla José Ana Fernandez Pérez 7, 38712, Breña Baja, Tenerife, Spain\\	
   \inst{10}INAF - Osservatorio Astrofisico di Catania, via S. Sofia 78, 95123 Catania, Italy\\    
   \inst{11}INAF - Osservatorio Astronomico di Brera, via E. Bianchi 46, 23807 Merate (LC), Italy\\
   \inst{12}Dipartimento di Fisica e Astronomia G. Galilei, Università di Padova, Vicolo dell’Osservatorio 2, 35122, Padova, Italy         
 	}		
 					
   \date{Received: \today; accepted XXX}
 
  \abstract
   {The distribution of exoplanets around low-mass stars is still not well
understood. Such stars, however, present an excellent opportunity of reaching
down to the rocky and habitable planet domains. The number of current
detections used for statistical purposes is still quite modest and different
surveys, using both photometry and precise radial velocities, are searching for
planets around M dwarfs.}
   {Our HARPS-N red dwarf exoplanet survey is aimed at the detection of new
planets around a sample of 78 selected stars, together with the subsequent
characterization of their activity properties. Here we investigate the survey
performance and strategy.}
   {From 2\,700 observed spectra, we compare the radial velocity determinations
of the HARPS-N DRS pipeline and the HARPS-TERRA code, we calculate the mean
activity jitter level, we evaluate the planet detection expectations, and we
address the general question of how to define the strategy of spectroscopic
surveys in order to be most efficient in the detection of planets.}
   {We find that the HARPS-TERRA radial velocities show less scatter and we
calculate a mean activity jitter of 2.3~m~s$^{-1}$ for our sample. For a
general radial velocity survey with limited observing time, the number of
observations per star is key for the detection efficiency. In the case of an
early M-type target sample, we conclude that approximately 50 observations per
star with exposure times of 900~s and precisions of about 1~ms$^{-1}$ maximizes the number of planet detections.}
   {}
   
   \keywords{Methods: statistical--Techniques: radial
velocities--Surveys--Stars: low-mass--planetary systems}

   \maketitle

\section{Introduction} \index{int} \label{int}

High-resolution time series spectroscopy employing the Doppler effect has been
very successful in the detection and confirmation of planets around bright
stars \citep{1995Natur.378..355M, 1996ApJ...464L.147M}. About 35~\% of the
planets discovered thus far have been found using this
method\footnote{\href{url}{http://www.exoplanets.eu}}. Modern spectrographs
like the High Accuracy Radial velocity Planet Searcher \citep[HARPS,
][]{2003Msngr.114...20M} are especially designed for this task
\citep{2008eic..work..375P} and actually a large majority of known sub-Neptune mass
planets were discovered by this instrument \citep{2004A&A...426L..19S,
2005A&A...443L..15B, 2007A&A...474..293B, 2006Natur.441..305L,
2006A&A...447..361U, 2007A&A...469L..43U, 2009A&A...493..639M}. More recent
instruments such as HARPS-N \citep{2012SPIE.8446E..1VC}, the recently commissioned CARMENES
\citep{2014SPIE.9147E..1FQ} or the upcoming ESPRESSO/VLT
\citep{2014AN....335....8P} are designed to extend the discovery space to
cooler and fainter objects.

The m~s$^{-1}$ precision level that these instruments can achieve allows the
detection of signals of small to large exoplanets (from super-Earths to
Jupiters) in relatively close-in orbits. However, the ultimate goal in the field
of exoplanetary astronomy is the finding of a true Earth twin, i.e. an Earth
mass planet orbiting a Sun-like star at a distance of about 1~AU. This is still
out of reach of current and upcoming instrumentation as the precision needed
is in the cm~s$^{-1}$ domain.  But, for the time being, the goal can be
generalized to finding rocky habitable exoplanets, preferably around nearby
stars. In this respect M dwarf hosts offer an interesting opportunity, although
they are not exempt of some challenges.

The ``fast track'' method of searching for rocky planets around low-mass M-type
stars is aided by {\em 1)} the larger radial velocity (RV) planetary signals
due to the reduced contrast in size with their host stars and {\em 2)} the
habitable zones with shorter orbital periods \cite[e.g.][]{1998A&A...338L..67D,
1998ApJ...505L.147M, 2004A&A...426L..19S, 2005ApJ...634..625R,
2006PASP..118.1685B}. On the other hand, M dwarfs have more efficient magnetic
dynamos than more massive stars, which results in a higher activity level
\citep{1984ApJ...279..763N, 2007AcA....57..149K} introducing additional RV
jitter. Also, they have been found to stay active for longer times during their
main sequence evolution \citep{1996AJ....112.2799H, 2008AJ....135..785W} which
makes the detection of an Earth-like planet around M stars more unlikely due to the increase of activity noise in the RV data and evolutionary effects. As we move from G to M dwarfs, the fraction of active
stars rises from a close to zero level to a peak of 90~\% for M7-M8 dwarfs \citep{2000AJ....120.1085G,
2013ApJ...769...37B, 2015ApJ...812....3W}.

Many surveys are in progress or planned to observe low-mass stars to detect
and characterize small rocky planets of less than 10~$M_{\oplus}$. The only
completed survey of this kind has been the HARPS M dwarf survey \cite[][hereafter Bo13]{2013A&A...549A.109B}, which includes observations of 102
pre-selected Southern M dwarfs. With an average of 20 observations per star
(obs/star), the authors provide orbital solutions for planets around their
targets and detect 9 planets. Better statistics were obtained by
\cite{2011arXiv1109.2497M}, who used 155 confirmed planets around 102 stars,
but without the focus on low mass stars. A significant step forward in the
statistical description of the planet population is expected from the just
started CARMENES survey, which will perform a 3-year intensive search for
planets (including habitable) around 300 Northern M dwarfs with high-resolution
optical and near-infrared Doppler spectroscopy \citep{2013hsa7.conf..842A}. The precision of those instruments is in the range of some 10~cm~s$^{-1}$ to 2~m~s$^{-1}$ depending on the brightness of the objects.

In general, the photometric {\it Kepler} mission \citep{2011ApJ...728..117B}
provides the best available statistics, although the transit technique does not
yield measurements of planetary masses ($M_{\rm P}$) but planetary radii
($R_{\rm P}$). The statistical studies published so far are not limited to
M-type stars or cover only a certain parameter space
\citep{2013ApJ...766...81F, 2013ApJ...767...95D, 2015ApJ...807...45D,
2013ApJ...767L...8K}.  \cite{2015ApJS..218...26S} for example used 163
planetary candidates with orbital periods($P_{\rm P}$) less than 200~days and
$R_{\rm P}<$13~$R_{\oplus}$ around 104 M dwarfs.

The studies published so far show that Neptune- and Earth-sized planets are
very abundant (around $\sim$40~\% of stars). The occurrence rates obtained for
Sun-like stars are sufficient to start putting together the picture of the
planet abundance as a function of $P_{\rm P}$ and minimum masses ($M_{\rm P}
\sin i$) or $R_{\rm P}$, although this is still not the case of M dwarf
planets.

In this work we describe the HArps-N red Dwarf Exoplanet Search (HADES)
program, where we monitor 78 bright Northern stars of early-M type (M0 to M3).
The observations began in August 2012 and we have recently reported a system of
two planets of $\sim$2 and $\sim$6~M$_{\oplus}$ around the M1 star GJ~3998
\citep[]{Affer2016}. The observations are ongoing and a number of
planetary candidates are being monitored. Once completed, our survey will
significantly strengthen the current statistics to determine planetary
frequencies and orbital parameters. Furthermore, our survey might reveal a
number of small, temperate, rocky, and even habitable planets worthy for
transit search and future atmospheric characterization.

Here we use the observational scheme of the HADES program (i.e. target
characteristics and time sampling), together with an underlying planet
population from state-of-the-art statistics to understand the yield of the
survey using the analysis tool presented by
\citet[][submitted]{GarciaPiquer2016}. The simulations allow us to obtain a
realistic estimate of the average level of uncorrelated magnetic activity
jitter in the sample. We also investigate possible observational strategies
aimed at maximizing the number of planet detections by running a trade-off
exercise between number of sample targets and number of observations per target
in the common case of a time-limited survey.

In Sect.~\ref{har}, we present our spectroscopic survey, introduce the dataset
and its treatment, and give an overview of the detection of periodic RV signals
due to magnetic activity and planetary companions. In Sect.~\ref{sta}, we
explain the simulations, compare them to our observations, and address the
important questions for the average noise level and observational strategy of a
survey around early M dwarfs. In Sect.~\ref{res} we draw the final conclusions
of this study.

\section{HARPS-N red dwarf exoplanet search} \index{har} \label{har}

The HADES program is a collaboration of the Spanish Institut de Ci\`encies de
l'Espai (ICE/CSIC) in Barcelona and the Instituto de Astrof\'isica de Canarias
(IAC) in Tenerife (EXOTEAM), and the Italian GAPS-M
project\footnote{\href{url}{http://www.oact.inaf.it/exoit/EXO-IT/Projects/Entries/2011/12/\\
27_GAPS.html}} \cite[Global Architecture of Planetary Systems -- M dwarfs, led
by G.~Micela, ][]{2013A&A...554A..28C, 2013A&A...554A..29D} including INAF
(Istituto Nazionale di Astrofisica) institutes in Catania, Milan, Naples,
Padova, Palermo, Rome, Torino and Trieste. In the following, we present the
program and its data set and explain the data treatment.

\subsection{Data acquisition and sample description} \index{dat} \label{dat}

We obtained optical spectra with the northern HARPS instrument \citep[HARPS-N,
][]{2012SPIE.8446E..1VC}, mounted at the 3.58~m Telescopio Nazionale Galileo
(TNG) in La Palma, Spain. Like its twin instrument in Chile, this is a fiber
fed, cross-dispersed echelle spectrograph with a spectral resolution of
115\,000, operating from 3\,800 to 6\,900~$\AA$. The stellar spectrum is split
into 66 diffraction orders or echelle apertures of 4\,096 pixels length over
the detector. It is designed to perform at high accuracy level using high
mechanical stability, avoiding spectral drifts with temperature and air
pressure adjustments, and operating in vacuum with a 1~mK stability. 
It is mounted at the Nasmyth B focus of the telescope and is fed by two fibres, 
one of which can be used for precise wavelength calibration by means of a Th-Ar emission lamp \citep{1996A&AS..119..373B}. 

The selection of the 78 stars monitored by the HADES program is described in detail in \citet{Affer2016}. We obtained 2674 spectra from August, 12$^{th}$ 2012 to March,
1$^{st}$ 2016, leading to an average of 34.3~obs/star, and adding up to a
total observation time of 891~h. In general, we decided not to include the simultaneous Th-Ar lamp calibration to avoid contamination in the science fibre due to the long integration times of 900~s and relative faintness of the targets. Because of this, we could not measure inter-night instrumental
drifts from our own observations. 
However, observations related to other
programs (using simultaneous reference light) were done by the GAPS team, and
we have around 5500 such drift values of the nights when our targets were
observed. Using those values we find the mean variation of the instrumental
drift in the data set to be of 1.00~m~s$^{-1}$ for the duration of the GAPS program.

	\begin{figure}[tbd]
		\resizebox{\hsize}{!}{\includegraphics[clip=true]{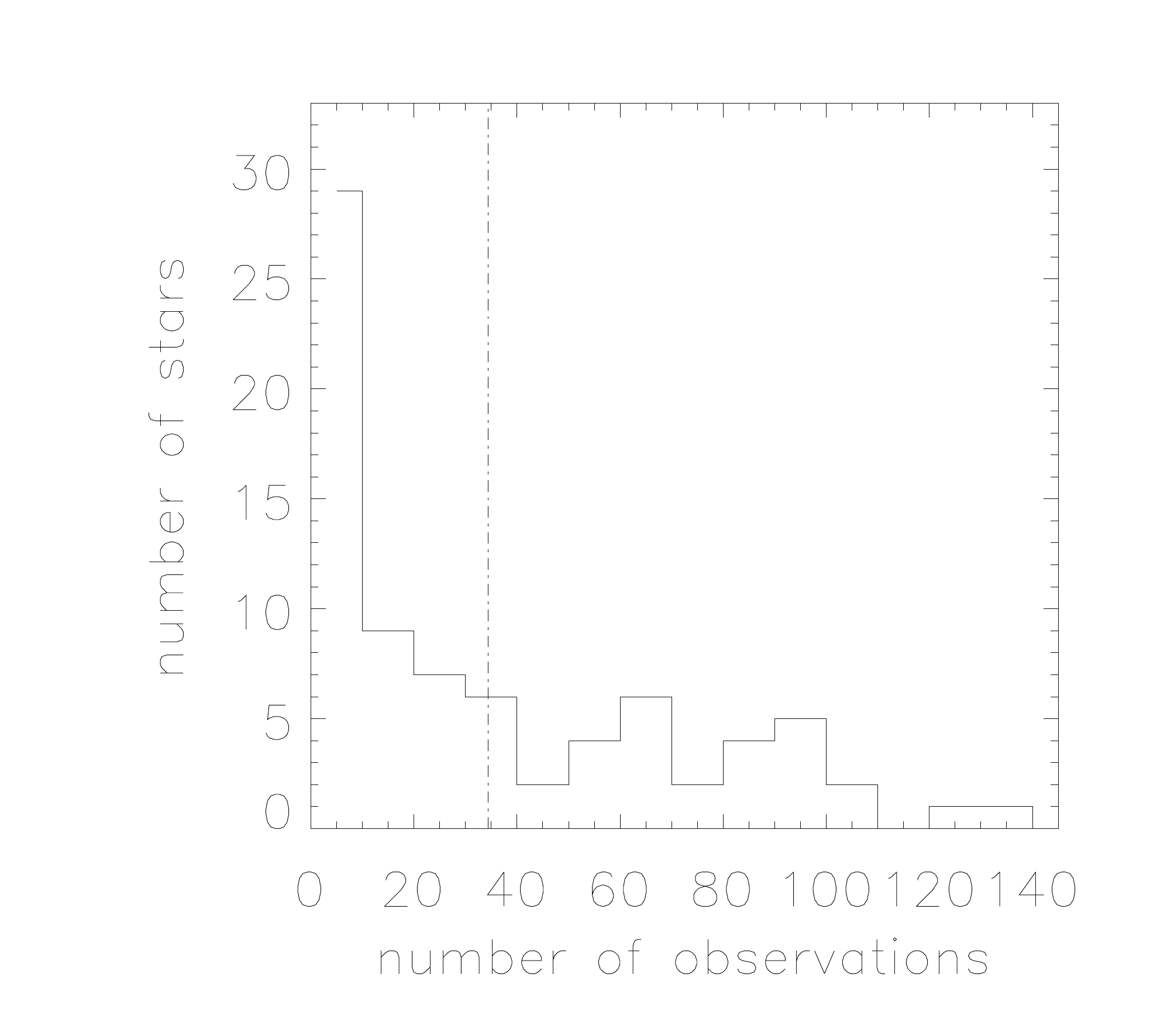}
			\includegraphics[clip=true]{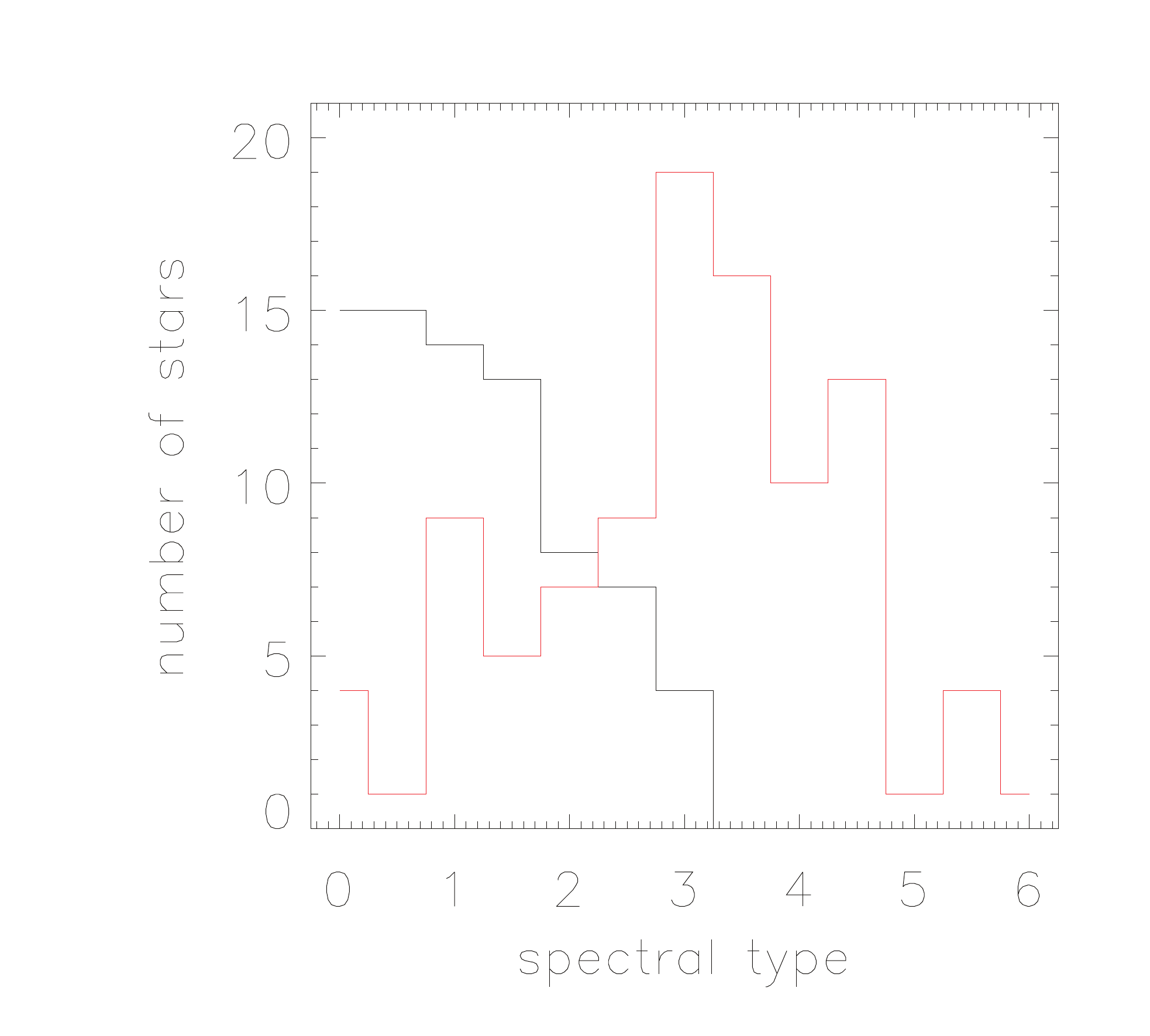}}
		\caption{\footnotesize \textbf{Left:} Distribution of the
number of obtained spectra for our targets as of March 2016. The dash-dotted
line indicates the average of 34.3 observations per star. \textbf{Right:}
Distribution of spectral types (M0 to M6) of our targets (black line) and Bo13
(red line).}
		\label{dis1}
	\end{figure}
	
	\begin{figure}[tbd]
		\resizebox{\hsize}{!}{\includegraphics[clip=true]{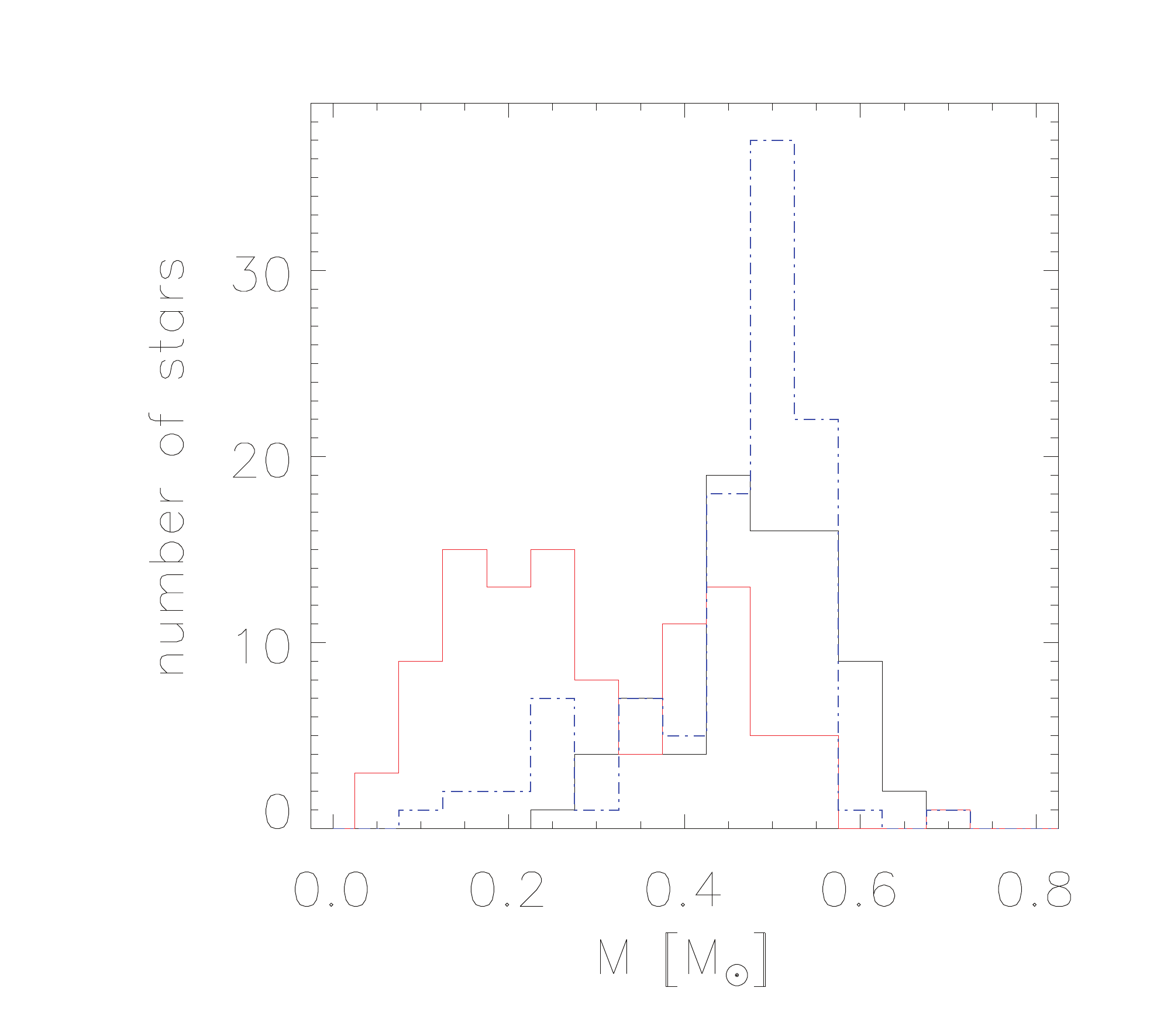} \includegraphics[clip=true]{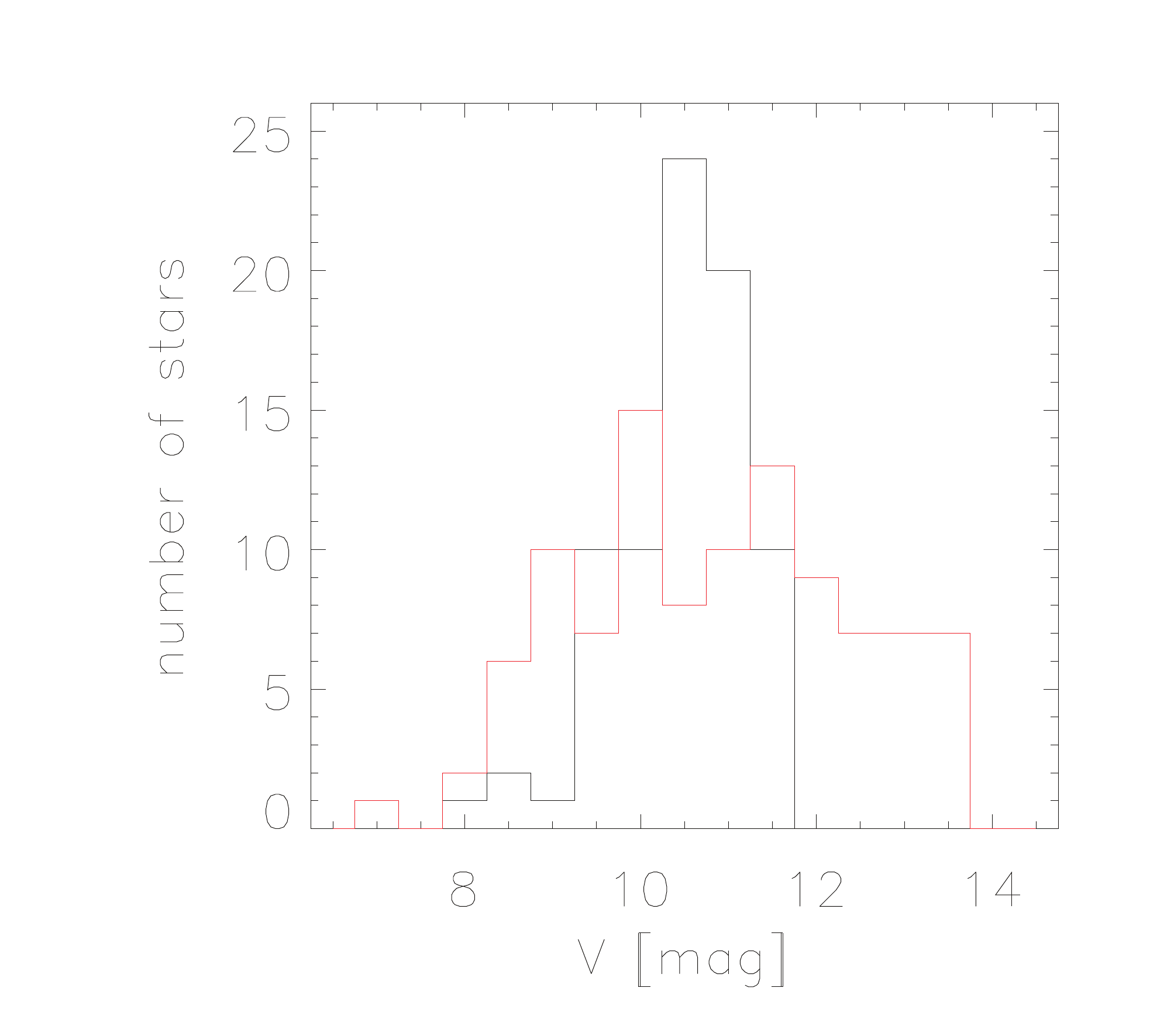}}
		\caption{\footnotesize Distribution of stellar masses (left)
and visual magnitudes (right) of our targets (black line), the known M dwarf data
from \citet[][ blue dash-dotted line]{2015ApJS..218...26S} and Bo13 (red
line).}
		\label{dis2}
	\end{figure}	
	
To estimate stellar properties, we apply the techniques introduced in
\cite{2015A&A...577A.132M}. Spectral types (SpT), effective temperatures, and
metallicities are obtained from calibrations of the ratios of sensitive line
features and their combinations. Those values are then used to determine
masses, radii, and surface gravities in an empirical way. The description of
the stellar parameters are provided with details in \citet{Maldonado2016}, whilst we show the relevant data for the present study
in Table~\ref{Tab1}. Data for targets without observations were collected from
the SIMBAD astronomical database \citep{2000A&AS..143....9W}, masses averaged
from the values of stars of same type from our sample.

In the left panel of Fig.~\ref{dis1} the distribution of number of spectra is
shown for our targets. 40 targets were observed more than 20 times, which is
the average number of the study by Bo13 and which we consider as a lower limit
necessary for the detection of planets. In the right panel, we see the
distribution of spectral types ranging from M0 to M3, with a peak at around M1. This narrow spectral type interval was purposely
chosen to minimize the number of parameters and derive a meaningful statistics.

In Fig.~\ref{dis2} we show the distributions of stellar masses and visual
magnitudes. They range from 0.30 to 0.69~$M_{\odot}$, with a peak at around
0.5~$M_{\odot}$, and 8.1 to 12.0~mag, respectively. The magnitudes of the
sample targets combined with the nominal selected integration time of 900~s
leads to a range of signal-to-noise ratios (S/N) at mid-wavelengths (spectral
order 46) of 23 to 163 (see Table~\ref{Tab1}). 
		
Our distribution of stellar masses is similar to the one used for the {\it
Kepler} stars by \cite{2015ApJS..218...26S}, while the sample of the HARPS
Doppler spectroscopy survey of Bo13 shows in average lower masses, later
spectral types and visual magnitudes of up to 14~mag. In this latter case, the
targets have an upper limit to the distance of 11~pc, whereas our objects show
an average of 19.2$\pm$8.5~pc and only 10 targets closer than such 11~pc limit.
The 9 detected planets by Bo13 orbit stars with  characteristics of M2.0$<$SpT$<$M3.0,
0.3$<$M$<$0.5~$M_{\odot}$, and 9.4$<$V$<$10.6~mag, covered by the HADES program.
		
\subsection{Data treatment} \index{datt} \label{datt}

The Data Reduction System pipeline \citep[DRS, ][]{2007A&A...468.1115L} of the
HARPS-N instrument reduces the data using the classical optimal extraction
method by \cite{1986PASP...98..609H} including bias and background subtraction
and flat-fielding and delivers cosmic ray-corrected, wavelength-calibrated
spectra. Furthermore, it calculates RVs using the cross-correlation function
(CCF) method \citep{1995IAUS..167..221Q, 1996A&AS..119..373B,
2002A&A...388..632P}. It calculates a contrast value by multiplying an observed
spectrum with a template mask. In our case, we use the mask of an M2 star, which
consists of around 9\,000 wavelength intervals of 820~m~s$^{-1}$ width (the HARPS-N pixel size) representing line features and their depths that are less affected by blending, and avoid regions of high telluric line
densities around 5\,320, 5\,930, 6\,300 and 6\,530~$\AA$ and which are weighted to take into account their expected S/N. The CCF is fitted by
the pipeline with a Gaussian function and the peak is the desired RV value.
All these reduction steps are done on each of the diffraction orders of the 2D
echelle spectra. The final measurements are then the flux-weighted mean of the
values measured across all orders. We re-reduced the observed data with YABI \citep{Hunter_researchopen, 2015A&A...578A..64B}. 
This tool is provided by the GAPS team and uses the same reduction steps as the 
DRS but giving the additional opportunity to change important values for the calculation of the CCF. 
YABI is mainly used to reduce the data in the same manner and to avoid small
changes made in the DRS during the years. An average RV uncertainty of 1.71~m~s$^{-1}$ is achieved for the YABI data set. The Geneva team
(F.~Pepe, priv. comm.) introduces an additional instrumental error of
0.6~m~s$^{-1}$ due to the uncertainty of the wavelength calibration. For more
details see \cite{2012SPIE.8446E..1VC} or the HARPS DRS
manual\footnote{\href{url}{http://www.eso.org/sci/facilities/lasilla/instruments/harps/tools/
\\ software.html}}.
 
We used a further approach to derive the RVs, namely the Java-based
Template-Enhanced Radial velocity Re-analysis Application \citep[TERRA,][]{2012ApJS..200...15A}. TERRA handles the full process of unpacking the
HARPS-N archive files in the DRS and YABI output. For each order it corrects
for the blaze function variability (i.e. flux) and, in a first run, does a
least-squares fit with each spectrum and the spectrum of highest S/N of each
target. With this information, a co-added template including all RV-corrected
input spectra of very high S/N is calculated. This is used as reference in a
second run of least-squares fitting to find the final RVs. TERRA offers the
opportunity of selecting the number of orders included in the calculation of
the RVs. For HARPS-N the usage from order 18 to 66 is recommended
(G.~Anglada-Escud\'e, priv. comm.) in order to minimize the RV root-mean-square
(rms) of the authors examples. This idea includes the assumption that the
smaller observed RV rms correspond to the smaller RV noise rms. The selected orders also correspond to the wavelength range of the M2 mask of the HARPS-N DRS. We obtain an
average uncertainty of 1.03~m~s$^{-1}$ with our dataset.

There are known shortcomings in the way in which the DRS and YABI deal with the
CCF resulting from the cross-correlation with the M2 mask. A symmetric
analytical function is fitted to the asymmetric CCF, and also the fit is
further complicated by the appearance of side-lobes caused by numerous lines included in the mask. Also, spectral type differences between the mask
and the target can affect the values of the CCF significantly, as our tests
with the StarSim simulator show \citep{2016A&A...586A.131H}. The TERRA
algorithm, on the other hand, introduces further error sources by using the
full spectrum including all blended lines and regions of higher telluric line
density.

\begin{figure}[tbd]
	\centering
	\includegraphics[clip=true, height=5.5cm]{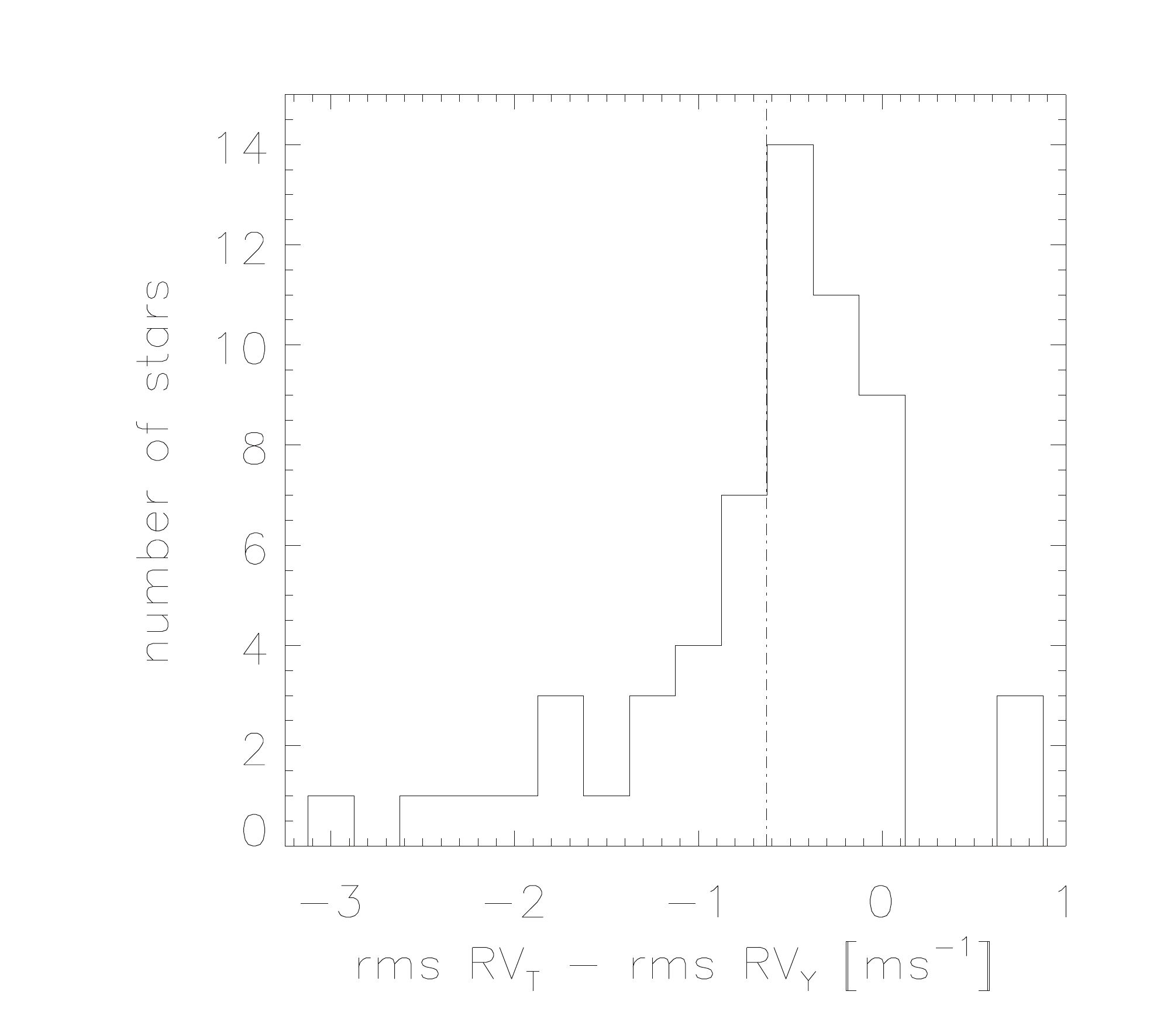}
	\caption{\footnotesize Distribution of RV rms differences from TERRA
and YABI for 62 targets. The mean value of -0.63~m~s$^{-1}$ is indicated by the dash-dotted line.
Note that 2 and 1 outliers are not shown on the left and right extreme ends of
the diagram, respectively (see Table~\ref{Tab1}). }
	\label{dis3}
\end{figure}

To verify which method delivers more precise results for our sample, we ran a
simple test of comparing the rms residuals of the RVs of each object using both
methods (see Fig.~\ref{dis3} and Table~\ref{Tab1}). The underlying idea is
that, as already mentioned, when considering a statistically significant
sample, the method that yields smaller rms of the RV measurements for the same
dataset should be preferred. Even in the presence of planets and activity
signals, more reliable measurements (i.e., more precise) should reflect into
smaller residuals. Also, it could be the case that the method yielding lower
rms values does so because of it being less sensitive to stellar activity
effects, and this also works in favour of improving the chances of
discovering exoplanet signals, which is the main goal.

The 62 stars with $>$1 observations show average rms values of 3.6 and
4.3~m~s$^{-1}$ and median rms of 2.9 and 3.4~m~s$^{-1}$ for TERRA and YABI
pipelines, respectively. The average difference between TERRA and YABI is
$-$0.63~m~s$^{-1}$ and there are only 13 targets showing larger rms values for
TERRA, which do not share any similar characteristics. Given the results of
this comparison, we adopt the RV data and associated error bars as derived by
TERRA.

\section{Simulations} \index{sta} \label{sta}

To address our main questions for the mean noise and activity level of early-M
dwarfs, the number of observations needed to detect a significant percentage of
planets, and the overall observational strategy, we use the code described by
\citet[][submitted]{GarciaPiquer2016}. This code was written to simulate the
outcome of the CARMENES observations after the application of the CARMENES
Scheduling Tool \cite[CAST, ][]{2014SPIE.9152E..21G}. With the input of stellar
properties such as the visual magnitude $V$, SpT, effective temperature, $M$,
the observational dates and the corresponding RV uncertainties, the code
generates simulated planets around the input stars and applies a search for
periodicity. 

\subsection{Adopted occurrence rates} \index{plas} \label{plas}

As further input to the code, we need to include distributions of various
planetary characteristics. Many studies have already been conducted to derive
such occurrence rates \citep{2002ApJ...566..463G, 2005PThPS.158...24M,
2005ESASP.560..833N, 2006Natur.443..534S, 2008PASP..120..531C,
2010ApJ...709..396B, 2010Sci...330..653H, 2010ApJ...717..878N,
2011ApJ...736...19B, 2012Natur.481..167C, 2012ApJS..201...15H,
2012A&A...541A.133Q, 2012ApJ...753..160W, 2013ApJ...764..105S,
2014ApJ...791...91C, 2015ApJ...809....8B} including different kinds of methods,
stellar environments and characteristics, but they are rarely focused on
low-mass stars. In the following we outline the parameter distributions
available and the ones used by \citet[][submitted]{GarciaPiquer2016} and our
work. 

For the orbital period the most significant and reliable statistics is provided
by the {\it Kepler} survey. Among the most comprehensive studies is that of
\cite{2013ApJ...766...81F}, who investigated 932 FGKM stars from
\cite{2013ApJS..204...24B} and analyzed around 180 planets with 0.8$< P_{\rm P}
<$418~days and 0.8$< R_{\rm P} <$22~$R_{\oplus}$.  \cite{2015ApJS..218...26S}
focused on M dwarfs and delivered distributions of $R_{\rm P}$ and $P_{\rm P}$
covering 0 to 13~$R_{\oplus}$ and 0 to 200~days, respectively.
\cite{2013ApJ...767...95D} and \cite{2015ApJ...807...45D} study 156 planets
detected in a sample of 2543 {\it Kepler} stars with T$<$4000~K with specific
interest in small planets in the habitable zone, and so the parameter intervals
are $P_{\rm P}$ from 0.5 to 200~days and $R_{\rm P}$ from 0.5 to 4~$R_{\oplus}$.
When correcting for detection biases, all these publications, together with
\cite{2013ApJ...767L...8K}, suggest a high occurrence rate of around 0.5 of
small planets around low-mass stars.

\begin{figure}[tbd]
	\resizebox{\hsize}{!}{\includegraphics[height=6.5cm]{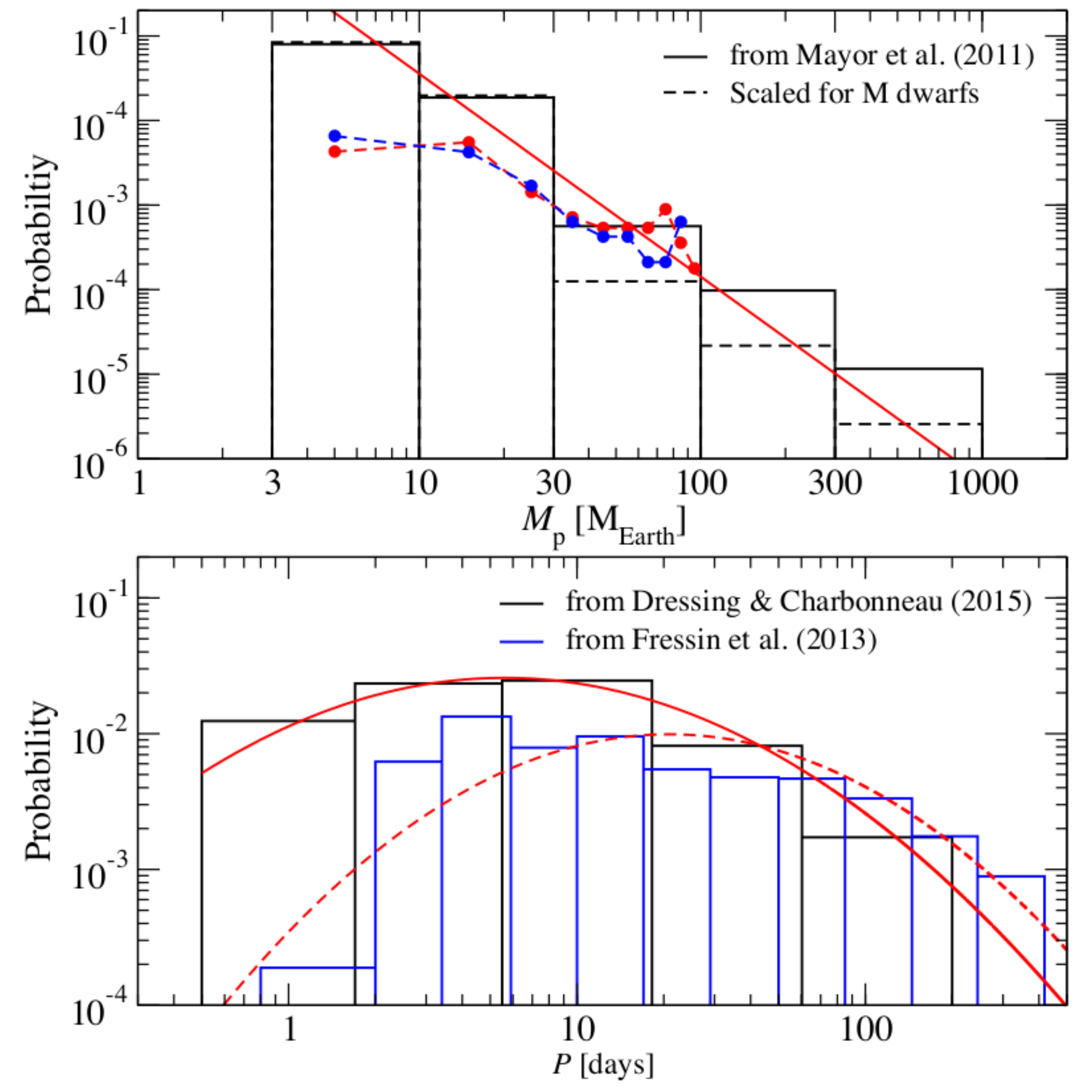}}
	\caption{\footnotesize \textbf{Top:} planet mass distribution assumed
for our simulations.  Solid bars show the probability distribution computed
from planet rates given by \cite{2011arXiv1109.2497M}. Dashed bars show the
scaled distribution for M dwarf host stars following the exoplanet statistics
from \cite{2013ApJ...767...95D}. Our fit to this scaled distribution is plotted
in red. These distributions are normalized between 3 and 1000~$M_{\oplus}$. We
show the distribution of minimum masses of the 155 planets from \citet[][dashed
red line and dots]{2011arXiv1109.2497M}, and the 81 theoretical solutions from
Bo13 (dashed blue line and dots).  \textbf{ Bottom:} orbital period
distributions corresponding to planet rates in
\citet[][blue]{2013ApJ...766...81F} and \citet[][black]{2015ApJ...807...45D}
used for planets with masses above and below 30~$M_{\oplus}$, respectively.
Solid and dashed red lines indicate the functional fits to these normalized
distributions.}
	\label{dis6}
\end{figure}

To describe the planet population we adopt a combination of these
distributions. For planets of $M_{\rm P}>$30~$M_{\oplus}$ (large Neptunes,
Jupiters, Giants) and 0.5$< P_{\rm P} <$418~days we follow
\cite{2013ApJ...766...81F}  and for $M_{\rm P}<$30~$M_{\oplus}$ (Earths,
Super-Earths and small Neptunes) and 0.5$< P_{\rm P} <$200~days we follow
\cite{2015ApJ...807...45D}. This division was also applied by a similar
simulation done by \cite{2015arXiv150603845S} for stars ranging from 0.15 to
0.78~$M_{\odot}$ and it takes into account the lower probability of planet
occurence on short periods. Additionally, in our simulation we use the
constraint of avoiding the generation of planets in the same system with period
ratios $<$1.3.  We use the following equation (see bottom panel of
Fig.~\ref{dis6}):
\begin{equation}
P(P_{\rm P}) dP_{\rm P}=D \cdot 10^{a+b \cdot \log(P_{\rm P})+c \cdot \log^{2}(P_{\rm P})} dP_{\rm P}.
\end{equation}
For planets with $M_{\rm P}>$30~$M_{\oplus}$ we use a=$-$4, b=2.205, c=$-$0.835, and
D=3.13 and for $M_{\rm P}<$30~$M_{\oplus}$ we use a=$-$2, b=0.954, c=$-$0.637, and
D=1.04, respectively.

For the distribution of $M_{\rm P}$, which is the relevant parameter in a
spectroscopic survey, we face the problem that the {\it Kepler} mission
statistics is based on measurements of $R_{\rm P}$, which is the observable
parameter resulting from transit modeling. Current observations show that the
relationship between $M_{\rm P}$ and $R_{\rm P}$ is very scattered and far from
being univocally defined. For this reason, we prefer to employ the statistical
distribution of planet masses coming from spectroscopic observations.
\cite{2011arXiv1109.2497M} use data of around 1200 inactive late-F to late-K
dwarfs resulting in the detection of 155 confirmed planets in 102 systems.
Bo13 provide the only significant statistics of planets around M dwarfs
considering $M_{\rm P}$ ranging from 1 to 10$^{4}$~$M_{\oplus}$ and periods
$P_{\rm P}$ ranging from 1 to 10$^{4}$~days from the detection of 9 planets in 102
targets. For our simulations we adopt the results of \cite{2011arXiv1109.2497M}, with a scaling factor for $M_{\rm P}>$30~$M_{\oplus}$ of 1.5\% following
\cite{2013ApJ...767...95D} in order to match the lower frequency of large planets around the low-mass stars. We reach up to 1\,000~$M_{\oplus}$ and extrapolate the fit below 3~$M_{\oplus}$ down to 1~$M_{\oplus}$. Although we expect that planets with lower masses are very abundant, their statistical distribution is not sufficiently well constrained and it is very unlikely for our simulations or the observational HADES program to detect them. We use the following equation (see top panel of Fig.~\ref{dis6}):
   	    \begin{equation}
   	    P(M_{\rm P}) dM_{\rm P}=1.40 \cdot M_{\rm P}^{-2.40} dM_{\rm P}.
   	    \end{equation}  
 
In this work we rather use the described functional fits instead of the tables
given in the publications for $M_{\rm P} $ and $P_{\rm P}$ in order to
approximately reproduce the expected behaviour of the probability
distributions. The solutions for M-type stars found by us are still unclear and
should be taken with some caution given the uncertainties of planet ratio
tables but we hope to be able to contribute to this interesting question once
the HADES program is finished.

\begin{figure}[tbd]
 	\resizebox{\hsize}{!}{\includegraphics[clip=true,height=7.5cm]{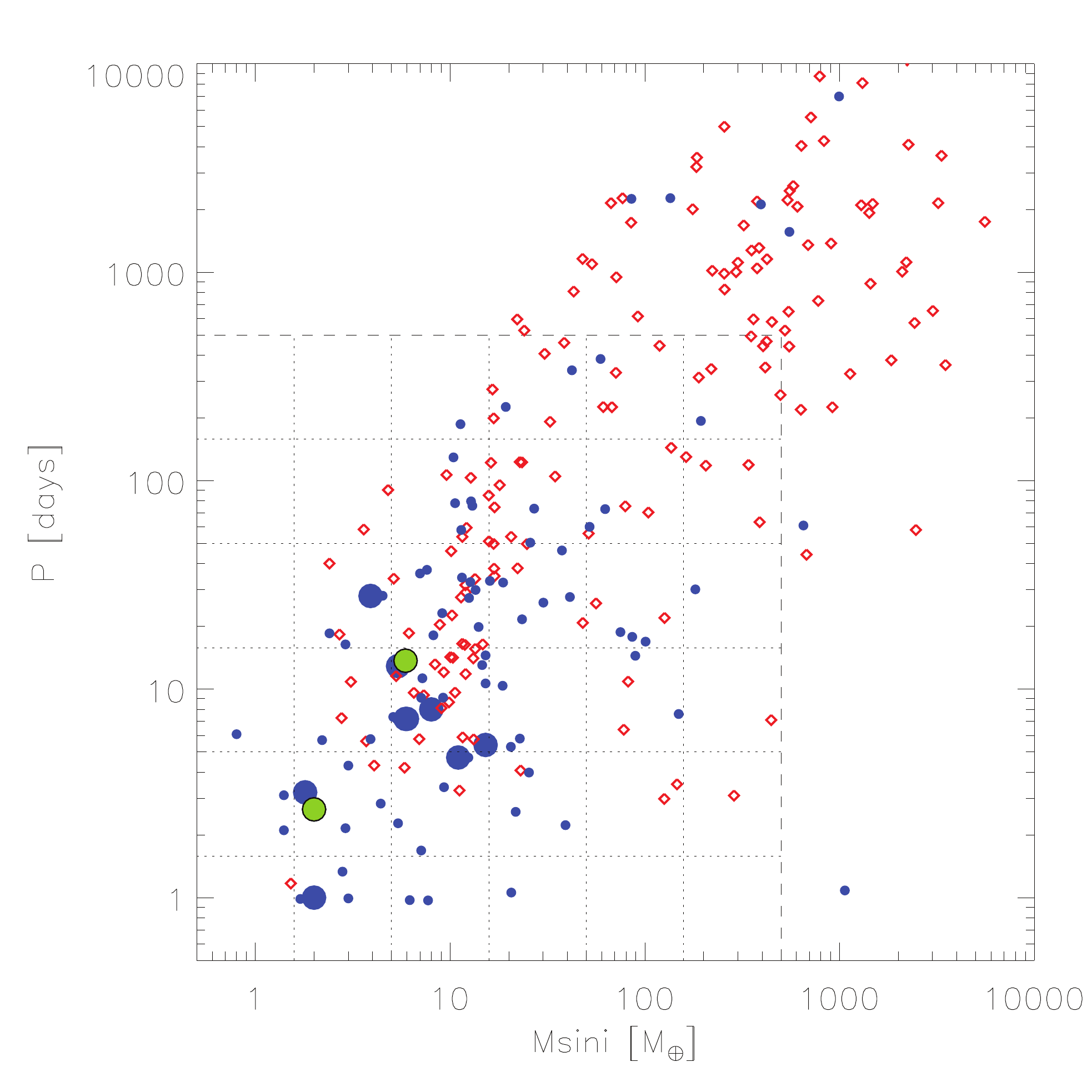}}
	\caption{\footnotesize Orbital period against minimum mass for 81 best
theoretical solutions for the targets of Bo13 (blue dots) and the 155
confirmed planets of \citet[][red diamonds]{2011arXiv1109.2497M}. The large blue and
green dots indicate the 9 planets detected by Bo13 and the 2 planets detected
by our group \citep{Affer2016}, respectively. The grid constructed out of the
dashed and dotted lines is for better comparison with Fig.~\ref{dis10}.}
	\label{dis5}
\end{figure}

In the comparison of the distributions of published $M_{\rm P} \sin i$ by
\cite{2011arXiv1109.2497M} and Bo13 and the models used, we see the
observational bias towards more massive planets, where the authors detect
basically all planets more massive than 20~M$_{\oplus}$, but miss the less
massive ones. We show a plot of $P_{\rm P}$ vs. $M_{\rm P} \sin i$ of the data
of the two mentioned publications in Fig.~\ref{dis5}. The values from the 155
confirmed planets by \cite{2011arXiv1109.2497M} are quite evenly distributed
over the plot except the region of low $M_{\rm P} \sin i$ and long $P_{\rm P}$
not technically accessible and the region of high $M_{\rm P} \sin i$ and low
$P_{\rm P}$ not populated by planets. The data of the 81 best theoretical
solutions for planets from Bo13 show few planets beyond 100~$M_{\oplus}$ and
100~days, such as their 9 detected planets.

Besides the distributions for $P_{\rm P}$ and $M_{\rm P}$, we use the
probability distributions of the sine wave of the inclination
(0$<i<$90$^{\circ}$) and the argument of periastron (0$< \omega
<$360$^{\circ}$), which we assume to be flat, and the eccentricities (0$<
\varepsilon_{\rm P} <$1) distributed following \cite{2013MNRAS.434L..51K} for
all stellar types with
    	\begin{equation}
    	P(\varepsilon_{\rm P}) d \varepsilon_{\rm P}=\frac{\Gamma(3.897)}{\Gamma(0.867) \cdot \Gamma(3.030)} \cdot \varepsilon_{\rm P}^{-0.133} \cdot (1-\varepsilon_{\rm P})^{2.030} d \varepsilon_{\rm P}.
    	\end{equation}

The multiplicity distribution of planets around M dwarfs can be taken from \cite{2015ApJ...807...45D} and the {\it Kepler} Object of Interest statistics for 1135 stars and show 58.8, 26.6, 8.6, 4.4, 1.4, 0.6, 0.6~\%  for stars with 1, 2, ..., 7 planets. Assuming that every star has a planet, we thereby distribute in average 1.70 planets per star, which is in the range of the values stated by \cite{2014MNRAS.443.2561G} and \cite{2015ApJ...807...45D}.

The construction of the parent distribution of planets is complex. This is
needed since any survey has its own properties and bias. We choose a
combination of various approaches and biases, since the real distribution is
unknown and since we do not consider it crucial for our goal to compute the
efficiency in detecting small planets and not only their absolute numbers.

\subsection{Simulation of observations} \index{mact} \label{mact}

To minimize numerical errors, we ran simulations considering 1\,000 scenarios,
i.e. different distributions of planets. On average, the realizations produce
128.1$\pm$8.3 planets around our 78 sample stars. Of them, 123.0$\pm$8.3
(96.0~\%) are terrestrial (i.e. Earths and Super-Earths with
$M_{P}<$10~$M_{\oplus}$), 3.0$\pm$1.6 (2.4~\%) are transiting planets ($\cos$~i$<
\frac{R_{S}}{a}$), and 27.3$\pm$4.9 (21.3~\%) are located in the habitable zone of
their respective host star. For the latter, we use the moist and maximum
greenhouse case for the inner and outer habitable zone following
\cite{2013ApJ...765..131K}.  
  
For each of the realizations, we use the underlying planet population to create
a simulated time series of RV measurements. We employ the real dates of our
observations and the RV uncertainties found by TERRA pipeline. The 54 stars
with more than 5 observations show mean and median observational RV rms of 4.0 
and 3.1~m~s$^{-1}$, respectively. The simulations of the RVs consider the contributions from the injected planets and the observational uncertainties, but lack the effects
introduced by magnetic activity or other additional noise sources. 

These values, however, are not completely representative of the magnetic
activity noise since part of the signal corresponds to correlated (periodic)
contribution that is usually removed during the detailed analysis by means
of, e.g., photometric time series or measurements of periodicities in activity
indicators such as H$\alpha$ and \ion{Ca}{II}. For a quantitative estimate of
this contribution to the RV rms we employ the identified periodicities
associated to stellar rotation, differential rotation, and their harmonics (Su\'arez Mascare\~no et al. 2016, in prep.), which reduce the RV rms of the stars
in average by 84\% (not considering the absolute RV rms). By also removing any long-term trends and applying a
3$\sigma$-clipping to the RV rms (rejecting TYC~2703-706-1 and NLTT~21156
because of their outlier values), we are left with a mean and median
observational RV rms of 3.0 and 2.8~m~s$^{-1}$, respectively, for the remaining
52 target stars.

\begin{figure}[tbd]
	\centering
	\includegraphics[clip=true,height=5.5cm]{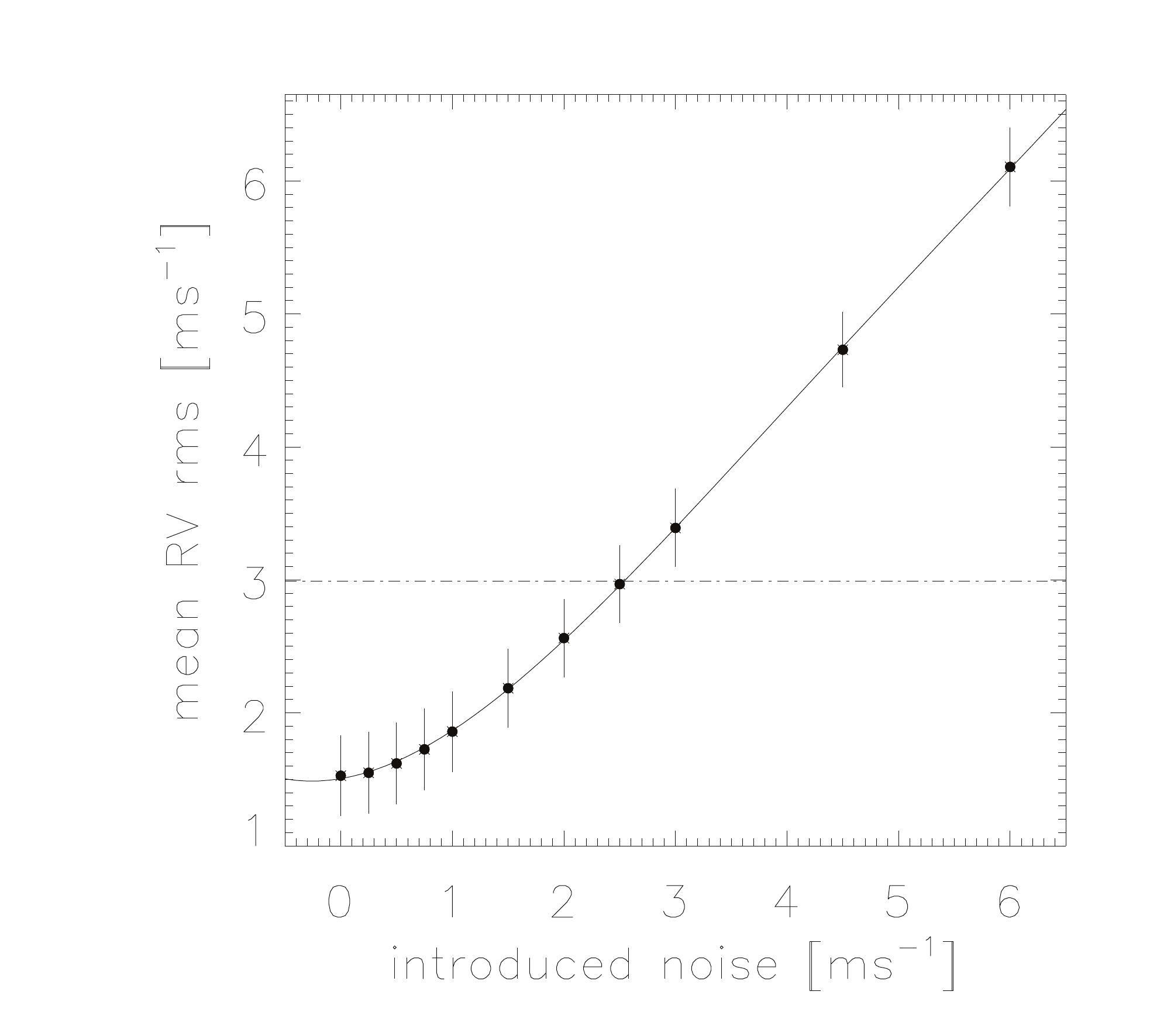}
	\caption{\footnotesize Relationship between the additional white noise
introduced quadratically in the simulated RV values and the resulting RV rms
for 52 target stars. The dots indicate the probed values, whereas the full
drawn line is the best polynomial fit. The dashed dotted line indicates the
mean observed value (without correlated activity jitter) of 3.0~m~s$^{-1}$.}
	\label{dis222}
\end{figure}	

To match this mean value, we searched for the correlation between a white noise
contribution added in quadrature and the resulting mean RV rms for our
simulation (see Fig.~\ref{dis222}). This was found to be
    	\begin{equation}
    	rms_{out}=a+b \cdot noise+c \cdot (noise)^{2} +d \cdot (noise)^{3} +e \cdot (noise)^{4}, \label{For1}
    	\end{equation}
    	with a = 1.506$\pm$0.241, b = 0.141$\pm$0.552, c=0.256$\pm$0.333 , d=$-$0.037$\pm$0.0.068 , and e=0.002$\pm$0.004.
    	
For our mean value of 3.0~m~s$^{-1}$, we find an additional noise of
2.6~m~s$^{-1}$. Note that this added noise is the combination of three sources: {\em 1)} the mean instrumental drift of 1.0~m~s$^{-1}$; {\em 2)} the
additional instrumental error of 0.6~m~s$^{-1}$ described above, and {\em 3)}
the uncorrelated magnetic activity (or RV) jitter \citep{2006ApJ...644L..75M,
2008ApJ...687L.103P}. If these noise contributions are treated as uncorrelated,
we obtain a mean value for the RV jitter associated to stellar activity of
2.3~m~s$^{-1}$. 

The simulated RVs so far consist of the observational uncertainties and the contribution of the simulated planets for each target. If we add quadratically this mean noise of 2.6~m~s$^{-1}$ to each target, we would end up with the same mean RV rms but not with the same RV rms distribution. And we would underestimate the number of targets with RV rms below the mean jitter value. Therefore, in the case that the simulated RV rms of a target is smaller than its observed one, we add quadratically a noise term to the simulations to match the observational RV rms (converted by Eq.~\ref{For1}). If the simulated RV rms of a target is larger than its observed one, we do not add any noise. We thereby create a similar simulated RV rms distribution as for our observations.

\begin{figure}[tbd]
 	\resizebox{\hsize}{!}{\includegraphics[clip=true, height=6.5cm]{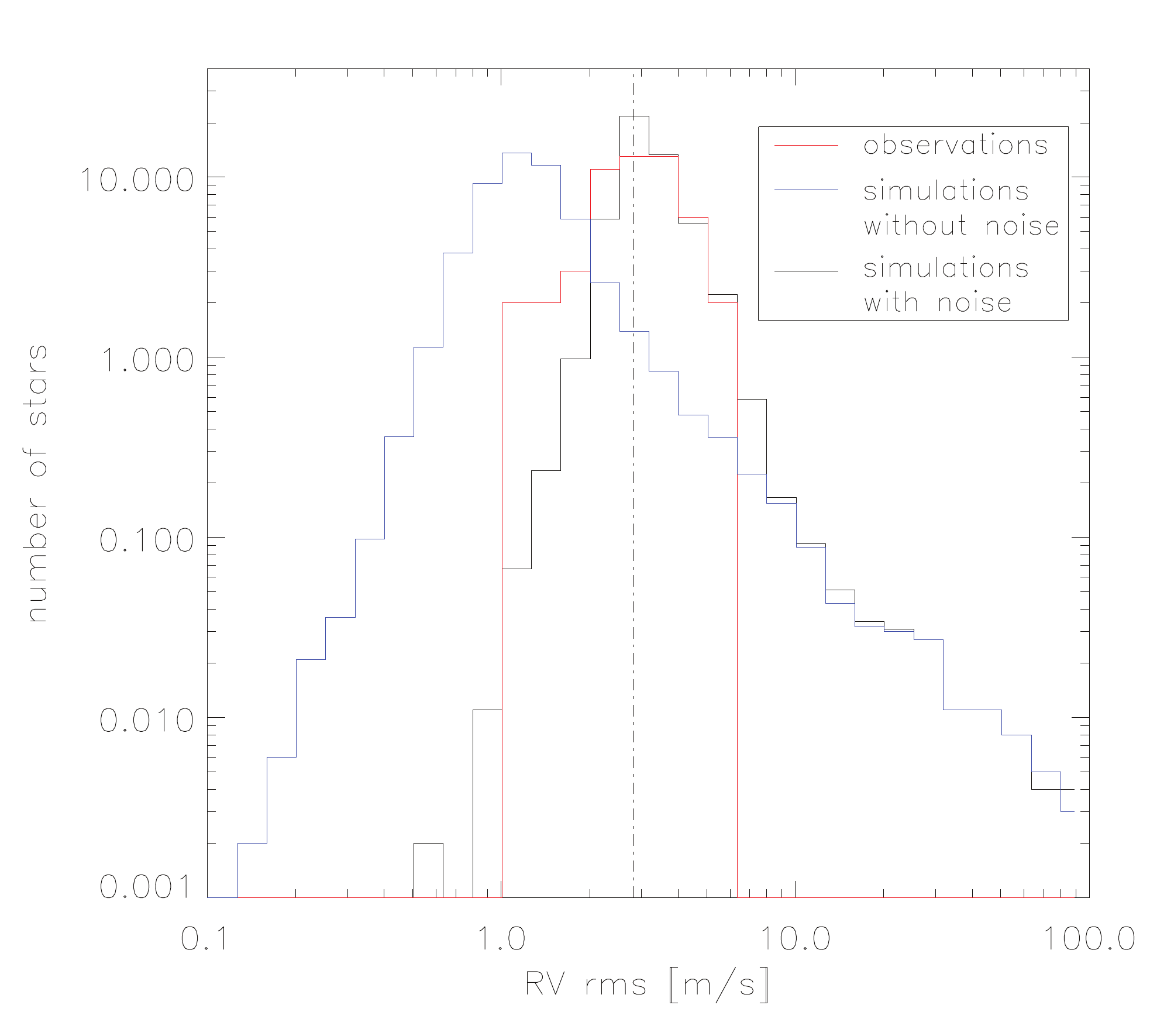}}
	\caption{\footnotesize Histogram of RV rms for 52 stars with more than
5 observations and a 3$\sigma$-clipping of the RV rms using observations (red),
initial (blue) and final (black) simulations. The initial RV rms distribution
does not include any activity jitter, whereas the final distribution includes
it by adding quadratically the RV rms as observed for each individual target.
For the simulations, we have uncertainties of the histogram values in the order of
$(0.92 N_{\rm S})^{0.5}$ and $(0.94 N_{\rm S})^{0.5}$ for the black and blue curve. respectively, with $N_{\rm S}$ being the number of stars. They are
not shown for reason of clarity.}
 	\label{dis7}
 \end{figure}

We show the RV rms
distributions of the 52 stars in Fig.~\ref{dis7}. The red curve indicates the
distribution as observed ranging from 1.1 to 6.0~m~s$^{-1}$, while the blue
curve shows the wide distribution simulated initially (note the logarithmic scale). The black curve
then is the final RV rms distribution after adding to each target in every
iteration the noise to reach its observed RV rms (without the correlated
activity jitter). The simulated targets with RV rms lower than 1.1~m~s$^{-1}$
are due to the Gaussian distribution applied and reach down to about
0.5~m~s$^{-1}$. The targets with RV rms larger than 6~m~s$^{-1}$ were
distributed by the simulation and not modified. By keeping the RV rms of every
target constant we introduce additional error sources: {\em 1)} the RV rms of
different spectral types are fixed, but we focus only on few different types
and their mean RV rms, ranging from 2.0 to 3.7~m~s$^{-1}$, do not differ
significantly from the overall one; {\em 2)} the RV rms of targets with
different numbers of observations are fixed, but the RV rms for targets with
40, 70, 90, and 100 observations, with 3.2, 3.2, 3.3, and 3.1~m~s$^{-1}$,
respectively, are in agreement with the overall mean RV rms. We note, that the
planetary candidates detected by the HADES program so far show host stars with RV rms of
5.6 and 4.2~m~s$^{-1}$ (including the planetary signals).

We then use the Generalized Lomb-Scargle periodogram \citep[GLS,][]{2009A&A...496..577Z}, which includes error-weighting and the consideration
of an RV offset, in order to search for the planets in our RV data. As
significance threshold, we use the power corresponding to a given False Alarm
Probability (FAP), which we estimate and calculate from the Horne number of
independent frequencies \citep{1986ApJ...302..757H}. An injected planet is detected, if {\em 1)} its period is connected to the highest peak of the GLS periodogram, {\em 2)} its FAP is below the usual threshold of 0.1~\%,  corresponding to a 99.9~\% probability level and {\em 3)} its frequency f=$\frac{1}{P}$ is located inside the frequency resolution $\delta f < \frac{1}{T}$ described by \cite{2009A&A...496..577Z}, where T is the observational time span of the target.
The subtraction of the most significant signal (assuming a Keplerian fit) and subsequent search for
additional signals is repeated until no relevant periodicity is left. In the
Keplerian fits and analysis of the residuals we do not consider linear trends
and adopt circular orbits, which is valid at least for short periods \citep[see
e.g.][]{2014ApJS..210...20M}. For further details of the simulating process, we
refer to \citet[][submitted]{GarciaPiquer2016}.

  \begin{figure}[tbd]
  	 	\centering  	
  	 	\includegraphics[clip=true, height=6.5cm]{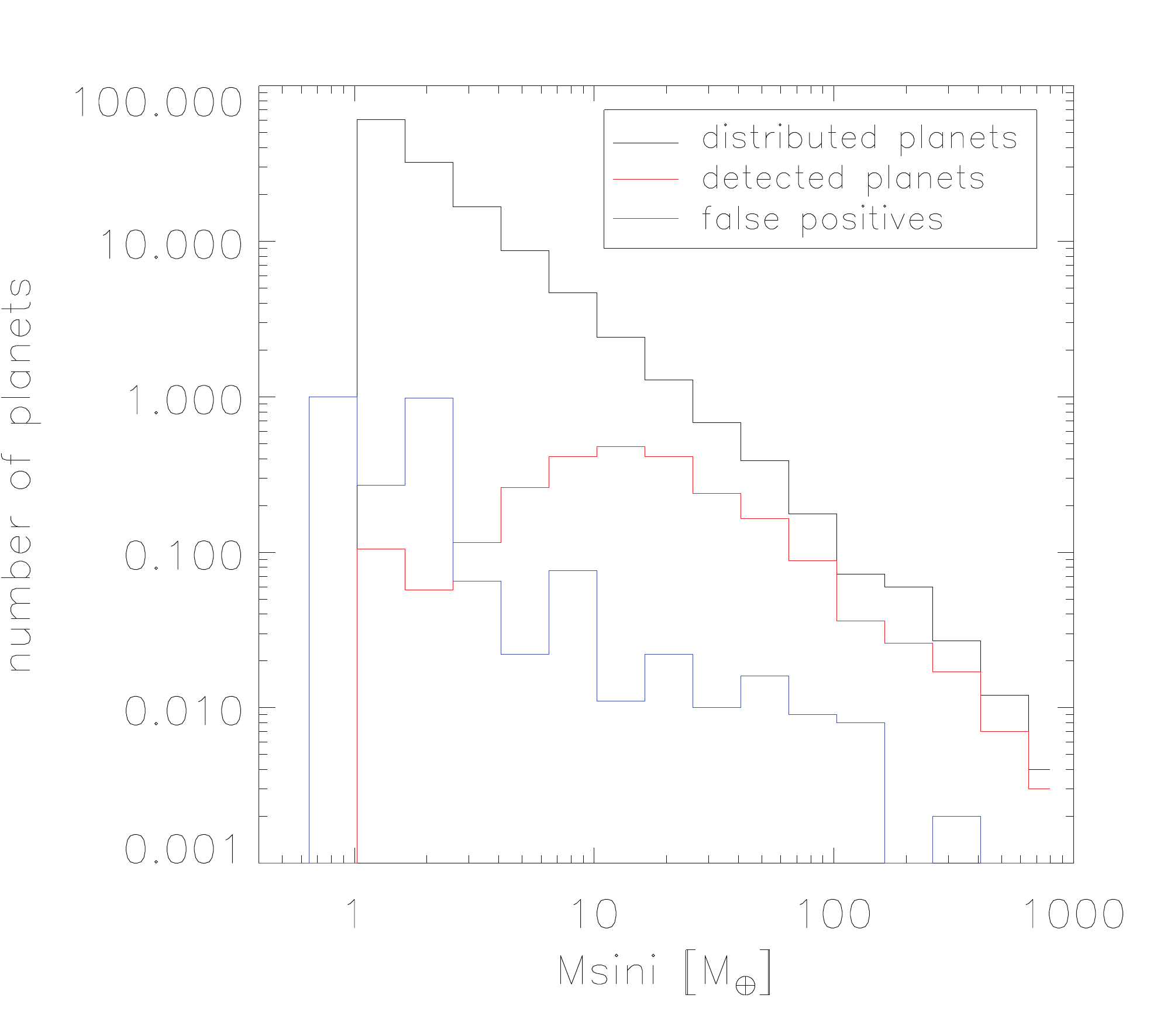}
  	\caption{\footnotesize 
		Histogram of minimum masses of the distributed planets around
our 78 targets (black line). The red curve indicates the respective
distribution of detected planets in our simulation comparable to the
observations and the blue line shows the false positives. Errors are in the
order of $(0.97 N_{\rm P})^{0.5}$ for all planets, $(1.00 N_{\rm P})^{0.5}$ for
the detected ones, and $(0.79 N_{\rm P})^{0.5}$ for the false
positives, respectively,. $N_{\rm P}$ is the number of planets.}
  	\label{dis8}
  \end{figure}	 

  \begin{figure}[tbd]
  	 	\centering  	
  	 	\includegraphics[clip=true, height=6.5cm]{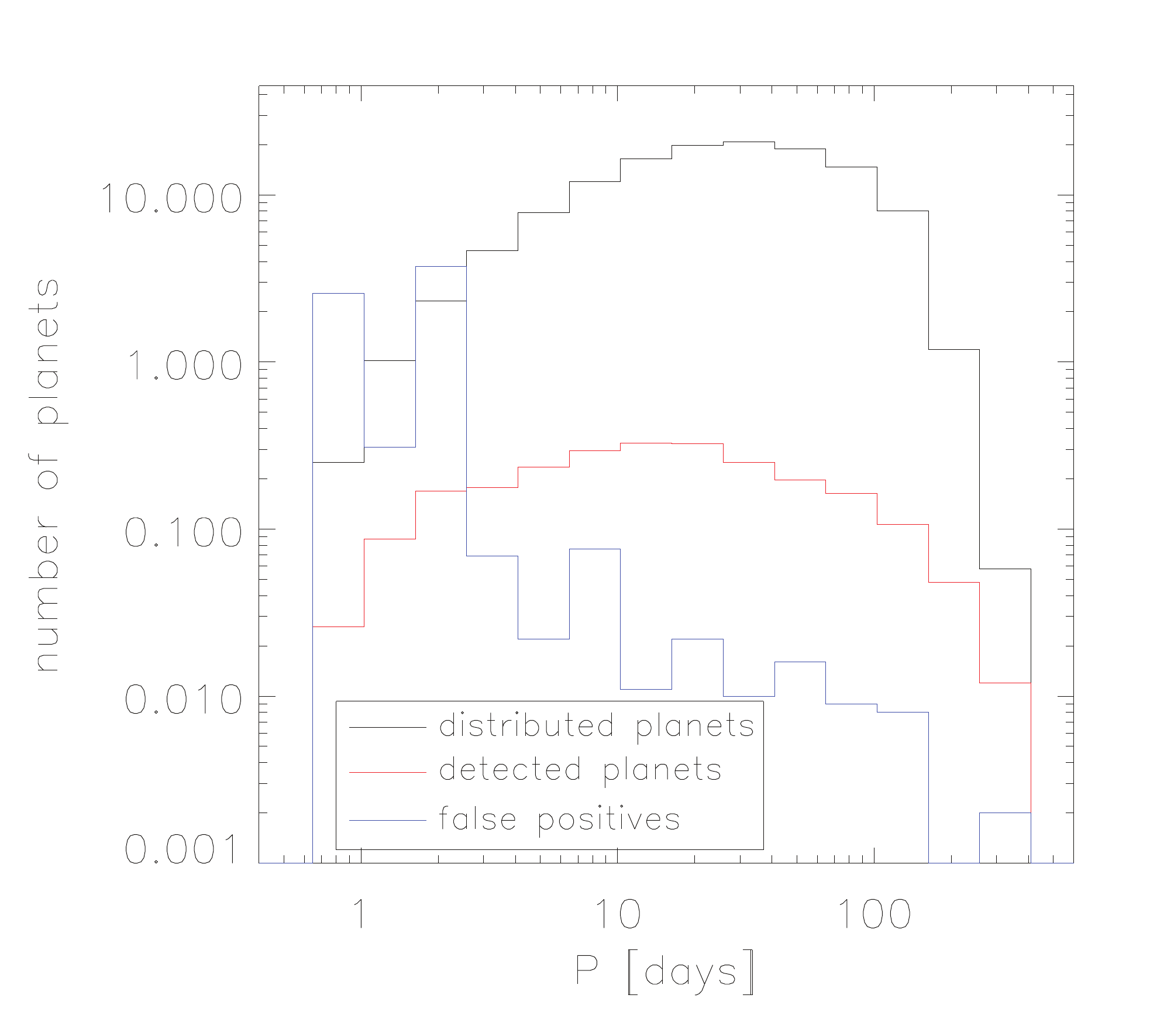}
  	\caption{\footnotesize 
		Histogram of orbital periods of distributed (black) and
detected (red) planets. Uncertainties are in the order of $(0.96 N_{\rm P})^{0.5}$ for
all planets and $(0.99 N_{\rm P})^{0.5}$ for the detected ones, respectively.
Also shown is the distribution of false positives (blue), with an uncertainty of
$(0.90 N_{\rm P})^{0.5}$.}
  	\label{dis8b}
  \end{figure}	           
          
\begin{figure}[tbd]
	\resizebox{\hsize}{!}{\includegraphics[clip=true, height=6.5cm]{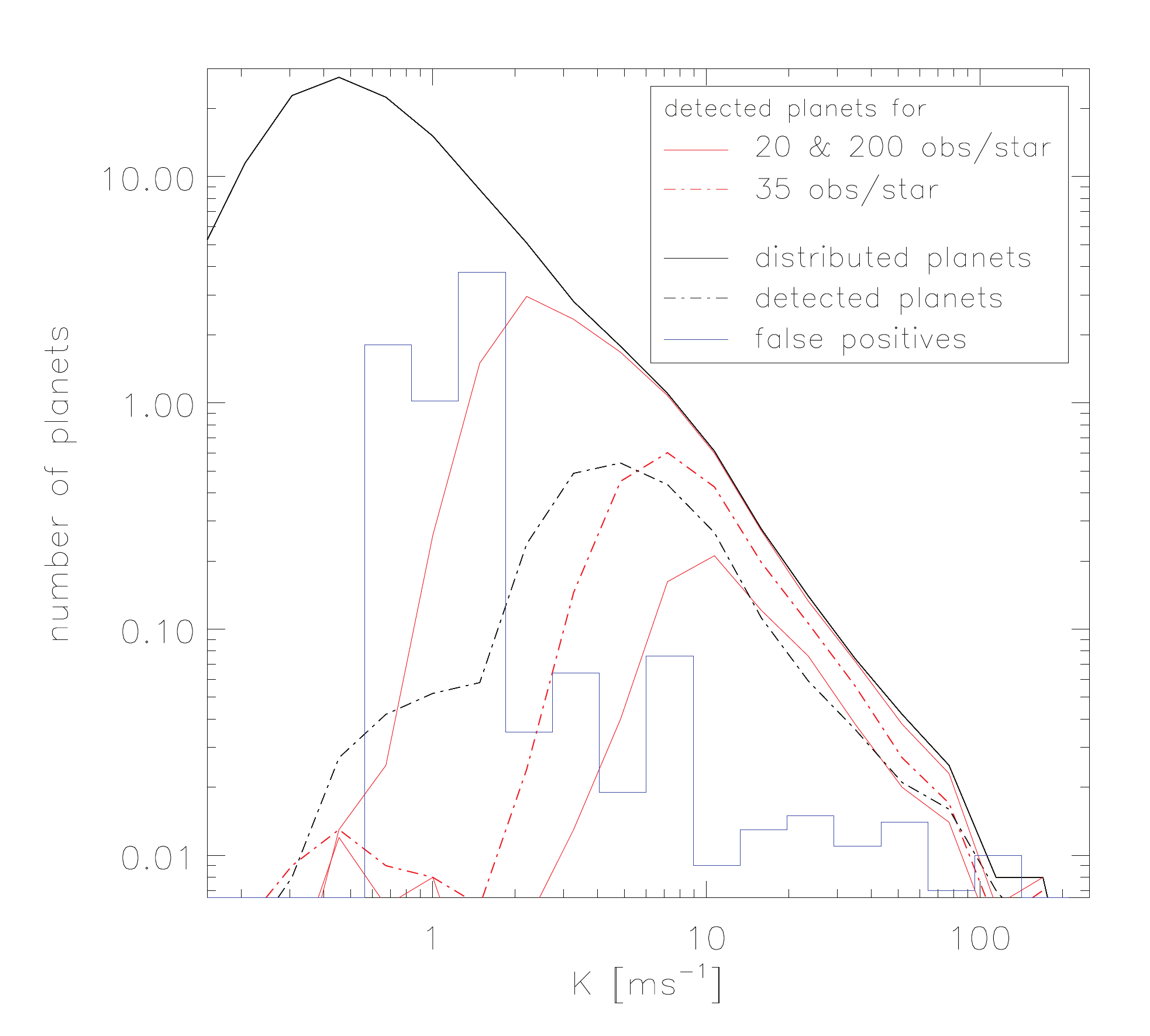}}
	\caption{\footnotesize Histogram of RV amplitudes $K$, where the
distribution for all simulated planets is shown by the solid black line with
uncertainties of $(0.97 N_{\rm P})^{0.5}$. We show the outcome of the simulations comparable to our observations
(uneven 34.3 observations per star) of Sec.~\ref{mact}, where the black
dash-dotted line indicates the planet detections and the blue histogram the
false positives with uncertainties of $(0.91 N_{\rm P})^{0.5}$. The red curves
indicate the distributions for detected planets for 20 (bottom), and 200 (top)
observations per star, the red dash-dotted line for even 35 observations per
star (see Sec.~\ref{nobs}). Uncertainties for all the curves of detected planets are in the order of $(0.98
N_{\rm P})^{0.5}$.} \label{dis9}
\end{figure}

The distributions of $M_{\rm P} \sin i$ and $P_{\rm P}$ are shown in
Figs.~\ref{dis8} and \ref{dis8b} with the distributed planets represented by
the black lines and the detected planets by the red lines. Also shown are the
distributions of false positive signals in blue. As already seen in the
comparison of theoretical models and observational data in Fig.\ref{dis6}, all
planets in our simulations more massive than 20~$M_{\oplus}$ are detected.
The highest detection rates we find for planets between 5 and 25~$M_{\oplus}$,
whereas for planets below 3~$M_{\oplus}$, the false positive rate is quite
high. The values here also go below our set limit of 1~$M_{\oplus}$, since the GLS detects them without considering our limit. In the case of the orbital periods, the
highest rate is found for 10--25~days and the detection curve obviously shows some bias
favouring shorter period planets. But in this range below 3~days, the false
positive signals are as numerous as generated planets, whereas they are
negligible for longer periods. If we integrate across all the parameters (i.e.,
area below the red lines in both Figs.~\ref{dis8} and \ref{dis8b}), our
simulations show that the observations of the HADES program analyzed here should
permit the detection of 2.4$\pm$1.5 planets (1.9\% detection rate). This is in agreement with the 3
planet detections announced by \cite{Affer2016} and Perger et al. (2016b, in prep.). Regarding the distribution of the simulated RV amplitudes,
Fig.~\ref{dis9} shows a maximum (black curve) at around 0.4~m~s$^{-1}$ and
values ranging from 0.01 to 100~m~s$^{-1}$. The decrease of the distribution at
low amplitudes is a consequence of the limit of 1~$M_{\oplus}$ applied to the
planets and not a physical effect. Planets with amplitudes K$>$5~m~s$^{-1}$ are
detected easily as seen by the curve of detected planets (black dash-dotted
line). The false-positive rate (in blue) rises very quickly for amplitudes
$<$2~m~s$^{-1}$, which is below the introduced mean RV jitter of
2.7~m~s$^{-1}$.

If we compare the outcome of the simulations with Bo13 and our own planetary
candidates, the mean minimum masses and mean periods found for all planets are
located in the described maxima of the curves. But it is notable, that the
detected planets by Bo13 show a mean amplitude of 4.7~m~s$^{-1}$, which is
nearly double of the one of our planets with 2.6~m~s$^{-1}$ and improves their
detectability. 

\subsection{Number of observations} \index{nobs} \label{nobs}

With the additional noise added to each individual target as described, we investigate the relation of the mean number
of observation per target and the efficiency in the detection of planets.  We
have run simulations for the 78 targets by using the same number of observations 
including 20, 35, 50, 70, 90, 120, 150, and 200~obs/star. We construct our own schedules cutting
out some already done observations and/or adding additional future observations
matching the usual distribution of observations of our teams (EXOTEAM and
GAPS-M) in the past semesters. Uncertainties are the ones observed with TERRA, or
Gaussian errors randomly distributed around the mean TERRA uncertainty of that
object (or objects of same spectral type, see Table~\ref{Tab1}), respectively.

\begin{figure}[tbd]
	\resizebox{\hsize}{!}{ \includegraphics[clip=true, height=6.5cm]{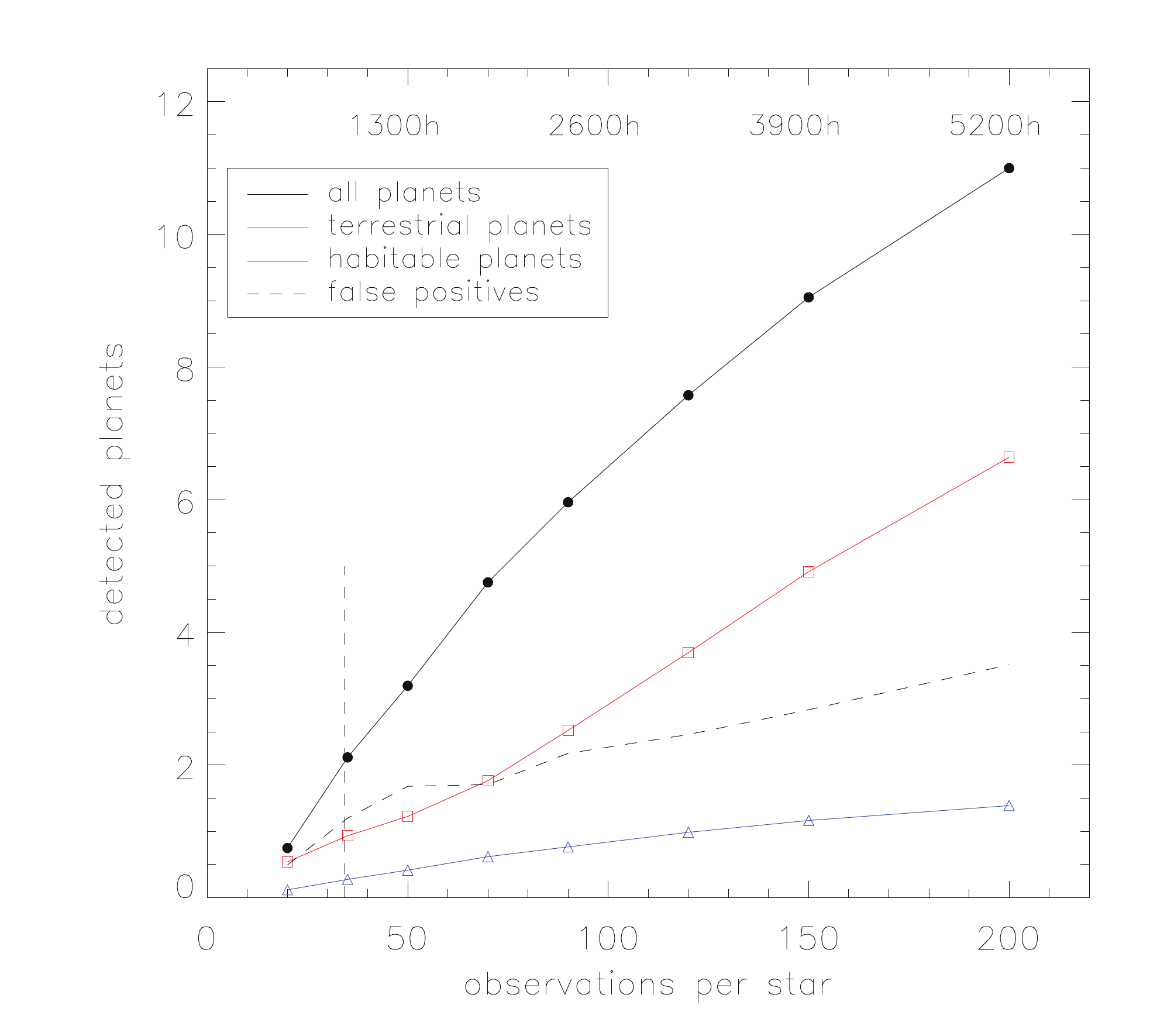}}
	\caption{\footnotesize Relation of the amount of all detected planets
		to the number of observations per star for the 78 sample stars (black dots, uncertainty $(0.92
		N_{\rm P})^{0.5}$), of the terrestrial planets (red squares, $(0.62 N_{\rm P})^{0.5}$), the habitable planets (blue triangles, $(0.95 N_{\rm P})^{0.5}$) and the false positives (black dashed line, $(1.17 N_{\rm P})^{0.5}$). We indicate on top the number of
		total observation times. With the 891~h total observation time and 34.3
		observations per star in average of our survey (vertical black dashed line), we detect around 2.1 planets resembling the detection rate for evenly
		distributed 35 observations per star.}
	\label{dis9b}
\end{figure}

In addition to the results explained in Sec.~\ref{mact}, Fig.~\ref{dis9} shows
the distribution of the RV semi-amplitudes K of the detected planets for
various fixed observations per star including 20, 35, and 200. Comparing the curve drawn by the simulations using an even 35~obs/star for all stars (dash-dotted red line) with the curve drawn by the
simulations of our observational case (dash-dotted black line) with unevenly distributed 34.3~obs/star in average, we observe a shift to smaller amplitudes of the curve,
equivalent to an improvement in the detection of planets with such
characteristics. This is obvious, since we are only able to detect those lower amplitude planets with a significantly higher number of observations. The increased detection of signals with K$<$1~m~s$^{-1}$ are the false positives included into the diagram.

The integrated number of planets of different types as a function of the number
of observations is shown in Fig.~\ref{dis9b}.  We reach a similar result as for
our actual observations with 34.3~obs/star and 891~h of total observation time
(2.4 planets detected) with an evenly distributed 35~obs/star. Spreading
observations among all targets is nearly as efficient as focusing on promising
targets. As expected, the curves follow approximately a square root function. We can compare the
results with the ones found for the CARMENES survey by
\citet[][submitted]{GarciaPiquer2016}, who use an average of around 100 to
110~obs/star and detect around 5\% of their planets adding an additional RV jitter of 3~m~s$^{-1}$. In our case, with a mean white noise contribution of
2.6~m~s$^{-1}$ we detect as well some 5\% of the distributed planets for the same amount of observations per star. Bo13 detect
around 6\% of the planets, if we use the same distribution rate as in our
study and put 160.8 planets around their 102 host stars. Although they
use only 20 HARPS spectra per target, they were able to pre-select and clean
their sample using a large number of lower resolution spectra from the ELODIE
(OHP, France) and FEROS (La Silla, Chile) spectrographs. And, as already
described, the planets detected by Bo13 show large mean RV amplitudes,
which do not seem to be present in our data.

 \begin{figure}[tbd]
 	\resizebox{\hsize}{!}{ \includegraphics[clip=true]{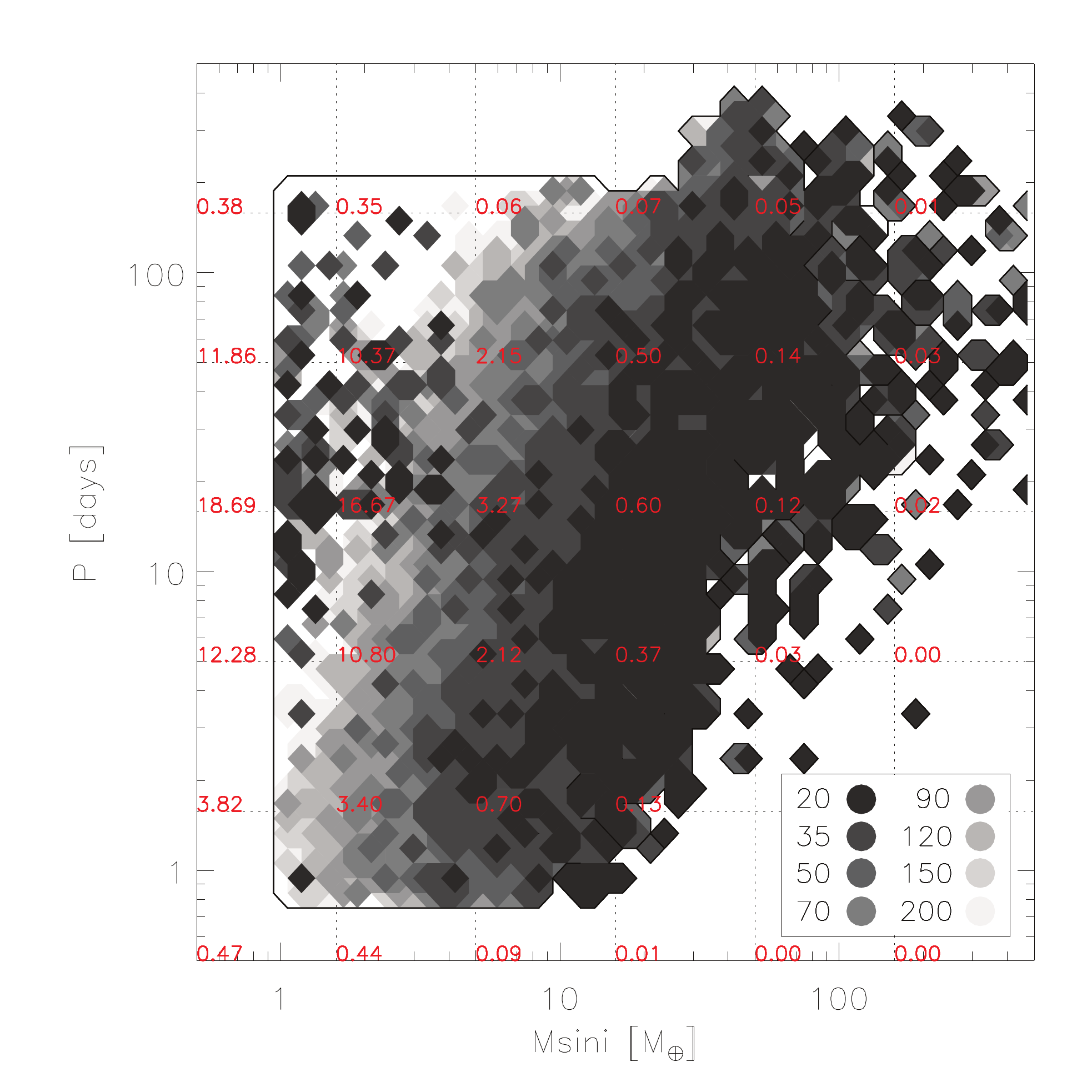} }
	\caption{\footnotesize Minimum masses vs. orbital periods of all the
simulated planets (black line) and their distribution in percent in various
subregions of the plot (red numbers) indicated by the dotted grid. From light
grey (200~obs/star) to black (20~obs/star), we show the distributions of every
detected planet of the 1\,000 simulations as indicated in the legend.}
 	\label{dis10}
 \end{figure}
 
In Fig.~\ref{dis10} we show $M_{\rm P} \sin i$ vs. $P_{\rm P}$ for all
distributed and detected planets. The black line shows the limits
of all planets distributed around the 78 sample stars. We also indicate in red
the percentage of planets located inside a logarithmic grid made out of the
dotted lines. From black to light grey we show the distributions of every
detected planet of the 1\,000 simulations from 20 to 200~obs/star as marked by
the color legend, respectively. As observable by the red marked numbers in the grid, the planets of our simulations are distributed around 1.5~$M_{\oplus}$ and 15~days, while the detected planets for each number of observations per star are
located along a line marked by the least massive planets. The distribution for 35~obs/star agrees well with the curves obtained by our observations in Figs.~\ref{dis8} and \ref{dis8b}. The false positive planetary detections are visible in the area of lower masses and longer periods.

Comparing the results to the data in Fig.~\ref{dis5}, many planets and
planetary candidates are located along the line of detection, but only few in
the center of distributed planets. As a rough comparison, most planets or
planetary candidates by \cite{2011arXiv1109.2497M} and Bo13 are detectable with 35 to 50~obs/star.

\subsection{Survey strategy} \index{rest} \label{rest}
 
We investigate further the observational
strategies by keeping constant the total duration of the survey and by varying
the number of target stars and the subsequent number of observations per
star. In short, we want to determine whether it is more efficient to observe
fewer targets many times or more targets a few times. 

\begin{figure}[tbd]
	\resizebox{\hsize}{!}{\includegraphics[clip=true, height=6.5cm]{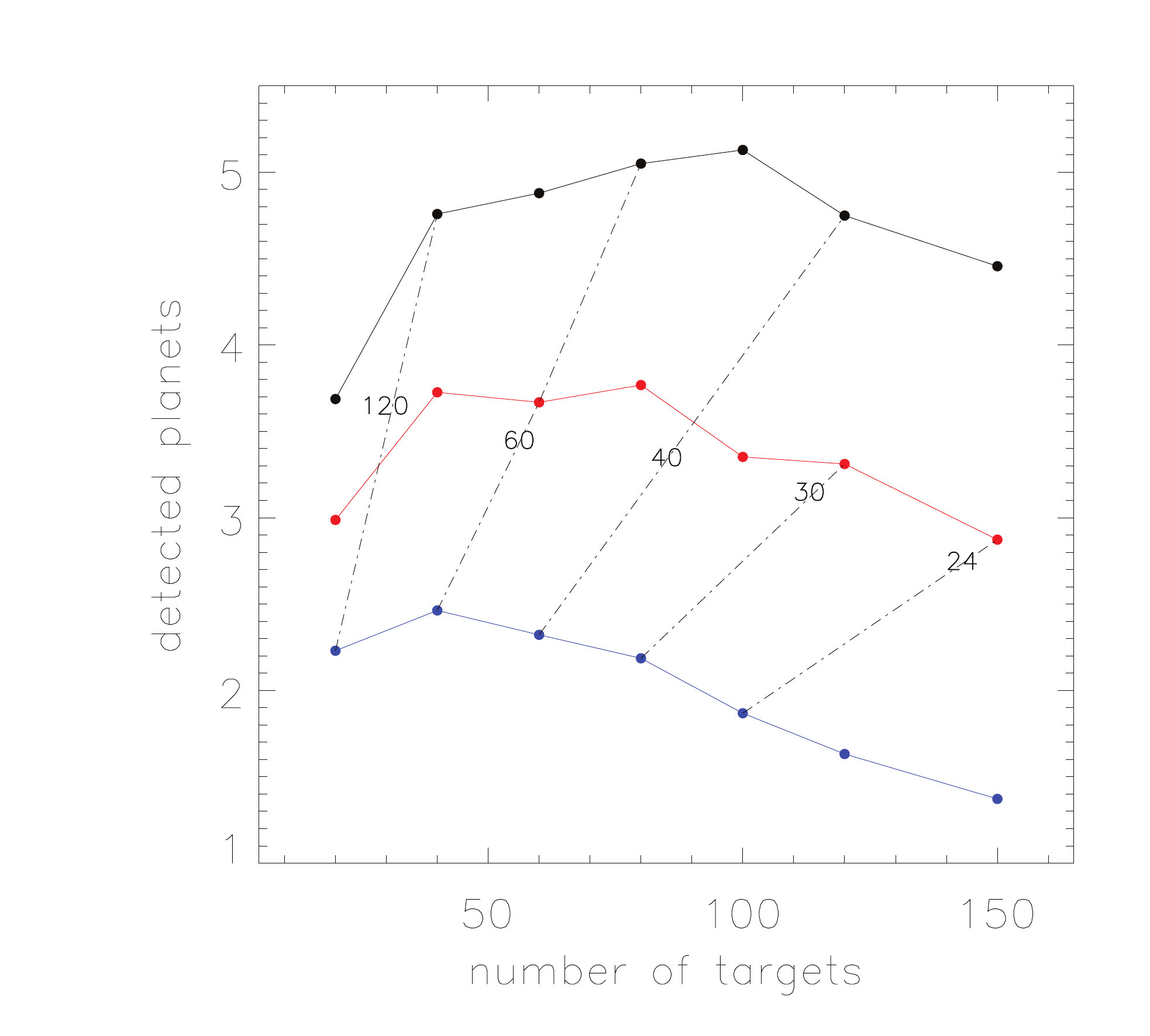}}
	\caption{\footnotesize Relationship between the number of detected
		planets of our simulations vs. the number of stars used for the survey. For a
		given total observation time of 800 (blue), 1\,200 (red) and 1\,600~h (black)
		we simulated 20 to 150 stars with 16 to 240 observations per star following
		Table~\ref{Tab2}. In the diagram, we connect points with the same number of
		observations per star using dash-dotted lines. Uncertainties are in the order of 1.4, 1.8, and 2.1 planets for the blue, red and black curve, respectively.}
	\label{dis11}
\end{figure}

We consider as a base 10 stars of our stellar sample selected in a way that the
spectral type distribution is similar to the original 78 star sample: two M0, M0.5, M1, and M1.5 stars and one M2 and M2.5 star (marked in Table~\ref{Tab1}). The sub-sample shows
a mean and median RV rms of 2.9 and 3.2~m~s$^{-1}$. We consider the similarity to the spectral type distribution and the overall RV rms as very important for the results of the simulations to be applicable to our entire sample. We do the simulations including
the sample various times using 20 to 150 stars as shown in Table~\ref{Tab2}. For the maximum number
of targets, we distribute 246.2$\pm$11.2 planets. We keep fixed the total
survey times of 800, 1\,200, and 1\,600~h, and calculate for each number of
stars the corresponding number of observations per target using the exposure time of 900~s. We employ the
observational uncertainties and the additional noise level used for each individual
target as explained in Sect.~\ref{mact}. We then compare the number of detected
planets as a function of the number of targets for a given total observation
time in Fig.~\ref{dis11}. 
 
\begin{table}[htb]
	\caption{\label{Tab2} Numbers of observations per target for different
numbers of target stars for three fixed total survey times of 800, 1\,200, and
1\,600~h.}
	\centering
	\begin{tabular}{c|ccc}
		\hline \hline
\#stars  & 800~h & 1200~h & 1600~h \\ \hline \hline
20 & 120 & 180 & 240 \\
40 & 60 & 90 & 120  \\
60 & 40 & 60 & 80  \\
80 & 30 & 45 & 60  \\
100 & 24 & 36 & 48  \\
120 & 20 & 30 & 40 \\
150 & 16 & 24 & 32  \\
		\hline \hline
	\end{tabular}
\end{table}

As expected, the total number of target stars that maximizes the number of
detected planets varies with the total survey time. The peak is located roughly
around 90--120 targets for 1\,600~h, and it progressively shifts to lower
number of targets as the total survey time decreases reaching a value of
30--50 for 800~h. We can estimate the maximum planet detection as a function of
the number of observations per star. We connect measurements of same number of
observations per target for different total observation times by the
dash-dotted line. In light of the plot, we can infer the position of the
maximum of the curves and we find that such value is around 40--60
observations per star for a survey of early-M stars distributed as the targets
of the HADES program. Thus, for 800, 1\,200, and 1\,600~h of observation (i.e., 100, 150
and 200 observing nights), the optimum number of targets with exposure times of
900~s to survey is, approximately, 46, 69 and 92 targets, respectively.
Surveys adopting a different strategy will suffer from diminished detection
efficiency.

\section{Conclusions} \index{res} \label{res}

We have presented the global analysis of the 2\,674 spectra of our HADES
program, in which we have been monitoring (and continue to observe) 78 M0 to
M3-type stars for the last four years with HARPS-N high-resolution Doppler
spectroscopy. We compare the two reduction pipelines DRS/YABI and TERRA and
find significant differences in the mean RV variations for our stars with 4.3
and 3.6~m~s$^{-1}$, respectively, and 79\% of the targets showing lower
variations for TERRA. Using the criterion that minimum rms value is associated
to a lower level of instrumental and stellar noise, we conclude that the TERRA
results are to be preferred as they seem to extract RVs with the best accuracy
and/or less affected by stellar magnetic activity. Following such a strategy,
careful measurements of RV and subsequent critical analysis has allowed the
discovery and recent announcement of a two-planet system of super-Earths by our
team \citep{Affer2016}. 

In this paper we also investigate the stellar noise properties of our early M
dwarf sample and, as a consequence, produce an estimate of the planet detection
rate that we compare with our results. We have performed a series of
simulations using state-of-the-art planet occurrence statistics applied to our
stellar sample and the actual observation times of our survey. Comparison
of the predicted RV variations with the real measurements yields a mean 
stellar noise level of 2.6~m~s$^{-1}$. Accounting for the typical HARPS
instrumental errors and drift values, we find an RV jitter level associated to
stellar activity of 2.3~m~s$^{-1}$. In our simulations we keep the RV rms of
each individual target constant since the distribution of the RV rms values
would be significantly broader in the case that we use the assumption of a mean
noise level for all stars. We predict that our survey should have been able to
detect significant signals from 2.4$\pm$1.5 planets, which is a rough estimate
but in good agreement with the announced discovery of 2 exoplanets so far. The
results show that planets with M$_{P}>$5~$M_{\oplus}$, K$>$2~m~s$^{-1}$ and
10$<P_{P} <$25~days are best detected which is confirmed by both the results
from Bo13 and the HADES program. For detections with masses, periods and amplitudes smaller
than the mentioned values, the false positive rate is significant.

With our simulations we also study the detection rate of planets while
increasing the number of obs/star from 20 to 200. This produces an increase in
the detection rate by an order of magnitude and with 95~obs/star, we are able to detect around 5\% of the distributed planets. We also note, that most of the available planet candidate signals
from Bo13 can be detected with around 35--50~obs/star, whereas with less
observations, the level of false positive detections is significantly higher. Those numbers are also necessary, to be able to detect planets of amplitudes around 2~m~s$^{-1}$. Results indicate that we best detect
planets with values around 10~days and $>$10~$M_{\oplus}$, while the simulated planets
are distributed around 1.5~$M_{\oplus}$ and 15~days. Although not
affecting the total number of detections, we can numerically demonstrate that
it is more efficient to concentrate on promising targets (those with
conspicuous signals) than to distribute observations equally amongst all
targets if searching for low-amplitude signals.

A further application of our simulation scheme is to assess the optimum number
of targets (and observations per targets) in the case of a time-limited survey,
which should be the most frequent scenario. Our analysis shows that the most
efficient strategy is to keep the average number of observations per target
constant and around a value of 50. This is, of course, for the case of an early
M-dwarf sample and exposure times of 900~s. Then, given an expected amount of observing time, the optimum
number of targets in the survey sample can be immediately calculated. 

We note that the rates of detected planets mentioned in this study may be
underestimated owing to the fact that we could not correct for the correlated
magnetic activity in all our targets. We include simulated host stars with
planetary companions reaching RV rms values beyond the 6~m~s$^{-1}$ observed by our HADES program.
Another note of caution should be added here concerning the still rather large
uncertainties in the exoplanet statistics around M dwarfs, which we aim to reduce with the HADES program. However, the planet
population seem to be sufficiently well established to reach conclusions on the
optimal survey design. We therefore suggest that surveys optimize their
strategy in light of our results to avoid having a diminished detection
efficiency.

\begin{acknowledgements}
The HARPS-N Project is a collaboration between the Astronomical Observatory of
the Geneva University (lead), the CfA in Cambridge, the Universities of St.
Andrews and Edinburgh, the Queens University of Belfast, and the TNG-INAF
Observatory. We made use of the SIMBAD database, operated at CDS, Strasbourg,
France.  Thanks to Guillem Anglada-Escud\'e for the TERRA reduction pipeline
and to the YABI project. GAPS acknowledges support from INAF through the
"Progetti Premiali” funding scheme of the Italian Ministry of Education,
University, and Research.  M. P., I. R., J. C. M., A. R., E. H, and M. L.
acknowledge support from the Spanish Ministry of Economy and Competitiveness
(MINECO) through grant ESP2014-57495-C2-2-R. J.I.G.H. acknowledges financial support from the Spanish Ministry of Economy 
and Competitiveness (MINECO) under the 2013 Ramón y Cajal program 
MINECO RYC-2013-14875, and A.S.M, J.I.G.H., and R.R.L. also acknowledge 
financial support from the Spanish ministry project MINECO AYA2014-56359-P.
\end{acknowledgements}

\bibliography{bibtex}{}
\bibliographystyle{aa}

\captionsetup{width=\textwidth}

\onecolumn
\scriptsize
\centering	

\begin{longtable}{l|cccc|ccccc|cc}

	\caption{\label{Tab1} Intrinsic and observational characteristics of
the 78 target stars of our sample sorted by number of observations (nobs). We show the
absolute RVs and their rms and the mean uncertainties dRV of every object for
TERRA (T) and YABI (Y) pipelines. $V$ magnitudes are from SIMBAD as are
spectral types of the sources flagged with (5). Their masses are the average
values of targets with same spectral type. The 10 stars flagged (1) are used in
Sect.~\ref{rest}, the 3 stars flagged (2) are the ones with larger rms
differences of TERRA and YABI mentioned in Fig.~\ref{dis3}. The targets flagged
(3) have TERRA uncertainties used for our simulations calculated as average
values of targets of similar magnitude. The targets flagged with (4) have
DRS/YABI uncertainties calculated the same way. The target marked with (6) has a companion discovered by \cite{2014ApJ...794...51H} with different data. Using HADES observations, the target marked with (7) has a companion published by \cite{Affer2016}.} \\

		\hline \hline
ID & nobs &  SpT & M & V & RV & RV$_{T}$ rms & dRV$_{T}$ & RV$_{Y}$ rms & dRV$_{Y}$ & S/N  & flag\\
&   &         &  [$M_{\odot}$] & [mag] & [kms$^{-1}$] & [ms$^{-1}$] & [ms$^{-1}$] & [ms$^{-1}$] & [ms$^{-1}$] & & \\
	\hline
	\endfirsthead
	\caption{continued.}\\
	
	\hline\hline
	
ID & nobs &  SpT & M & V & RV & RV$_{T}$ rms & dRV$_{T}$ & RV$_{Y}$ rms & dRV$_{Y}$ & S/N  & flag\\
&   &         &  [$M_{\odot}$] & [mag] & [kms$^{-1}$] & [ms$^{-1}$] & [ms$^{-1}$] & [ms$^{-1}$] & [ms$^{-1}$] & & \\ 
	\hline
	\endhead
	\hline
	\endfoot		
		
GJ~3998 & 137 & M1 & 0.50$\pm$0.05 & 10.85 & $-$44.81 & 4.24 & 1.17 & 4.99 & 1.82 & 42.6 & (7) \\
GJ~4306 & 119 & M1 & 0.53$\pm$0.05 & 10.62 & $-$31.72 & 2.85 & 0.98 & 3.17 & 1.60 & 52.9 & -- \\
GJ~16 & 106 & M1.5 & 0.48$\pm$0.05 & 10.90 & $-$14.84 & 2.25 & 1.06 & 2.63 & 1.59 & 43.9 & -- \\
GJ~694.2 & 102 & M0.5 & 0.55$\pm$0.06 & 10.72 & 4.62 & 3.73 & 1.26 & 5.37 & 2.08 & 46.8 & (1) \\
GJ~3942 & 97 & M0 & 0.63$\pm$0.07 & 10.19 & $-$18.71 & 5.61 & 1.10 & 5.87 & 1.87 & 59.2 & \\
GJ~625 & 97 & M2 & 0.30$\pm$0.07 & 10.13 & $-$12.85 & 2.68 & 0.78 & 2.54 & 1.01 & 64.7 & (1) \\
GJ~49 & 94 & M1.5 & 0.55$\pm$0.05 & 9.57 & $-$5.78 & 5.47 & 0.77 & 5.92 & 1.08 & 81.6 & -- \\
GJ~2 & 93 & M1 & 0.51$\pm$0.05 & 9.93 & $-$0.04 & 3.69 & 0.79 & 4.03 & 1.07 & 72.7 & (1) \\
GJ~119A & 90 & M1 & 0.55$\pm$0.05 & 10.73 & 76.84 & 2.49 & 0.94 & 2.89 & 1.53 & 55.0 & -- \\
GJ~15A & 88 & M1 & 0.38$\pm$0.05 & 8.10 & 12.00 & 2.29 & 0.49 & 2.25 & 0.48 & 163.1 & (6) \\
GJ~4057 & 84 & M0 & 0.59$\pm$0.07 & 10.78 & 0.86 & 2.96 & 1.17 & 3.46 & 2.16 & 47.3 & (1) \\
GJ~21 & 81 & M1 & 0.53$\pm$0.05 & 10.51 & $-$2.78 & 4.57 & 1.19 & 5.16 & 1.85 & 51.9 & -- \\
GJ~740 & 81 & M0.5 & 0.58$\pm$0.06 & 9.23 & 10.62 & 4.15 & 0.69 & 4.55 & 0.97 & 97.1 & (1) \\
GJ~720A & 77 & M0.5 & 0.57$\pm$0.06 & 9.85 & $-$31.31 & 3.60 & 0.79 & 3.80 & 1.26 & 71.9 & -- \\
GJ~412A & 70 & M0.5 & 0.38$\pm$0.05 & 8.71 & 69.09 & 2.14 & 0.61 & 2.30 & 0.81 & 115.7 & -- \\
TYC~2703-706-1 & 68 & M0.5 & 0.64$\pm$0.06 & 11.87 & $-$21.71 & 28.58 & 2.61 & 25.79 & 5.26 & 29.5 & (2) \\
GJ~47 & 66 & M2 & 0.36$\pm$0.06 & 10.83 & 7.76 & 2.86 & 0.96 & 3.18 & 1.42 & 43.3 & -- \\
GJ~9440 & 66 & M1.5 & 0.51$\pm$0.05 & 10.62 & $-$11.50 & 3.70 & 1.14 & 2.79 & 1.70 & 44.6 & (1) \\
GJ~156.1A & 63 & M1.5 & 0.55$\pm$0.05 & 10.86 & $-$20.32 & 2.68 & 1.08 & 3.19 & 1.71 & 44.4 & (1) \\
GJ~3997 & 63 & M0 & 0.49$\pm$0.05 & 10.36 & $-$20.57 & 3.19 & 1.07 & 4.28 & 1.79 & 54.7 & (1) \\
GJ~552 & 60 & M2 & 0.47$\pm$0.05 & 10.69 & 8.04 & 2.76 & 0.97 & 3.00 & 1.41 & 43.9 & -- \\
GJ~162 & 58 & M1 & 0.50$\pm$0.05 & 10.18 & 35.15 & 3.42 & 0.90 & 4.71 & 1.44 & 63.9 & (1) \\
GJ~9689 & 57 & M0.5 & 0.57$\pm$0.06 & 11.30 & $-$67.70 & 4.65 & 1.39 & 5.75 & 2.85 & 35.2 & -- \\
GJ~150.1B & 50 & M1 & 0.51$\pm$0.05 & 10.83 & 34.86 & 4.64 & 1.15 & 4.47 & 1.73 & 47.2 & -- \\
GJ~521A & 50 & M1.5 & 0.47$\pm$0.05 & 10.24 & $-$65.16 & 2.24 & 0.85 & 2.87 & 1.22 & 54.3 & -- \\
GJ~685 & 44 & M0.5 & 0.55$\pm$0.06 & 9.98 & $-$14.70 & 6.49 & 0.87 & 6.93 & 1.42 & 67.2 & -- \\
GJ~184 & 43 & M0.5 & 0.54$\pm$0.05 & 9.96 & 65.95 & 2.59 & 0.89 & 2.79 & 1.43 & 67.9 & -- \\
GJ~26 & 39 & M2.5 & 0.37$\pm$0.07 & 11.06 & $-$0.17 & 2.82 & 1.03 & 3.07 & 1.54 & 38.7 & (1) \\
GJ~3822 & 39 & M0.5 & 0.56$\pm$0.06 & 10.65 & $-$7.89 & 6.24 & 1.27 & 6.89 & 2.28 & 44.3 & -- \\
GJ~408 & 35 & M2.5 & 0.35$\pm$0.07 & 10.01 & 3.34 & 2.18 & 0.74 & 2.38 & 1.03 & 56.4 & -- \\
GJ~9404 & 33 & M0.5 & 0.62$\pm$0.07 & 10.62 & $-$0.50 & 3.83 & 0.98 & 4.01 & 1.96 & 51.2 & -- \\
NLTT~21156 & 33 & M2 & 0.50$\pm$0.05 & 11.23 & 14.08 & 16.94 & 2.27 & 18.64 & 3.30 & 25.3 & -- \\
GJ~399 & 30 & M2.5 & 0.55$\pm$0.06 & 11.30 & 3.43 & 3.25 & 1.41 & 4.01 & 2.31 & 28.9 & -- \\
GJ~414B & 29 & M2 & 0.50$\pm$0.05 & 9.98 & $-$15.16 & 2.00 & 0.74 & 2.13 & 1.17 & 57.9 & -- \\
GJ~548A & 29 & M0 & 0.63$\pm$0.08 & 9.76 & 9.63 & 4.30 & 0.69 & 4.09 & 1.46 & 71.2 & -- \\
GJ~1074 & 26 & M0.5 & 0.52$\pm$0.05 & 10.97 & 17.30 & 3.43 & 1.28 & 3.52 & 2.28 & 42.5 & -- \\
GJ~606 & 26 & M1.5 & 0.46$\pm$0.05 & 10.50 & $-$16.95 & 3.54 & 1.08 & 3.32 & 1.83 & 44.1 & -- \\
GJ~793 & 25 & M3 & 0.33$\pm$0.08 & 10.58 & 10.78 & 2.02 & 0.82 & 1.98 & 1.18 & 49.2 & -- \\
GJ~450 & 23 & M1.5 & 0.45$\pm$0.05 & 9.74 & 0.47 & 3.58 & 0.94 & 4.27 & 1.48 & 61.5 & -- \\
GJ~70 & 21 & M2.5 & 0.37$\pm$0.06 & 10.93 & $-$25.71 & 2.76 & 0.94 & 3.71 & 1.47 & 39.4 & -- \\
GJ~3649 & 17 & M1.5 & 0.50$\pm$0.05 & 10.76 & 31.74 & 2.10 & 1.15 & 2.73 & 2.04 & 43.2 & -- \\
GJ~2128 & 16 & M2.5 & 0.34$\pm$0.06 & 11.49 & $-$30.40 & 2.28 & 1.03 & 2.27 & 1.92 & 32.0 & -- \\
GJ~731 & 14 & M0 & 0.57$\pm$0.06 & 10.15 & $-$14.34 & 2.21 & 0.69 & 2.59 & 1.85 & 68.1 & -- \\
BPM~96441 & 13 & M0 & 0.66$\pm$0.08 & 11.84 & 6.42 & 3.56 & 1.41 & 5.58 & 3.55 & 34.9 & -- \\
GJ~4092 & 13 & M0.5 & 0.62$\pm$0.07 & 10.86 & $-$82.80 & 3.08 & 0.95 & 4.21 & 2.31 & 45.5 & -- \\
GJ~476 & 12 & M3 & 0.38$\pm$0.07 & 11.42 & 33.47 & 1.98 & 0.99 & 2.94 & 1.79 & 31.5 & -- \\
NLTT~53166 & 11 & M0 & 0.58$\pm$0.06 & 11.11 & 9.36 & 1.43 & 1.12 & 3.12 & 2.88 & 40.2 & -- \\
StKM~1-650 & 10 & M0.5 & 0.61$\pm$0.07 & 11.96 & $-$18.15 & 3.50 & 1.80 & 5.33 & 4.11 & 27.3 & -- \\
V*BR Psc & 10 & M1.5 & 0.37$\pm$0.06 & 8.99 & $-$70.95 & 2.67 & 0.71 & 2.86 & 0.91 & 97.2 & -- \\
GJ~3117A & 8 & M2.5 & 0.43$\pm$0.06 & 11.38 & $-$12.60 & 3.10 & 1.02 & 3.56 & 1.93 & 33.7 & -- \\
GJ~272 & 7 & M1 & 0.50$\pm$0.05 & 10.56 & $-$30.99 & 4.48 & 1.36 & 8.48 & 2.60 & 39.4 & (2) \\
GJ~3352 & 7 & M0.5 & 0.56$\pm$0.06 & 11.07 & $-$71.69 & 2.82 & 0.78 & 2.80 & 2.18 & 41.9 & -- \\
TYC~3379-1077-1 & 7 & M0 & 0.69$\pm$0.08 & 11.56 & 29.04 & 2.28 & 2.04 & 9.79 & 5.55 & 23.6 & (2) \\
GJ~671 & 6 & M2.5 & 0.31$\pm$0.09 & 11.38 & $-$19.32 & 1.146 & 0.61 & 1.03 & 1.35 & 39.5 & -- \\
2MASS~J22353504+3712131 & 4 & K7.5 & 0.62$\pm$0.07 & 11.93 & 5.76 & 3.32 & 0.93 & 3.44 & 4.30 & 32.6 & -- \\
GJ~1030 & 4 & M2 & 0.50$\pm$0.05 & 11.43 & 17.12 & 0.83 & 0.57 & 1.32 & 1.94 & 36.7 & -- \\
GJ~895 & 4 & M1.5 & 0.54$\pm$0.05 & 10.03 & $-$33.00 & 1.25 & 0.48 & 0.41 & 1.11 & 70.9 & -- \\
GJ~119B & 3 & M3 & 0.47$\pm$0.06 & 10.99 & 76.03 & 2.35 & 0.24 & 5.16 & 2.15 & 29.6 & -- \\
GJ~3126 & 2 & M3 & 0.45$\pm$0.07 & 11.13 & $-$83.94 & 0.02 & 0.01 & 0.25 & 1.17 & 44.0 & -- \\
GJ~686 & 2 & M1 & 0.42$\pm$0.05 & 9.62 & $-$9.33 & 0.85 & 0.02 & 1.77 & 1.02 & 72.6 & -- \\
TYC~3720-426-1 & 2 & M0 & 0.62 & 11.34 & $-$3.29 & 2.18 & 0.15 & 1.28 & 4.92 & 38.0 & (5) \\
TYC~743-1836-1 & 2 & M0 & 0.62$\pm$0.06 & 10.93 & 39.90 & 0.24 & 0.03 & 2.57 & 2.74 & 44.5 & -- \\
GJ~3014 & 1 & M1.5 & 0.48$\pm$0.05 & 11.12 & $-$15.39 & -- & 0.95 & -- & 3.01 & 24.5 & (3) \\
GJ~3186 & 1 & M1 & 0.53$\pm$0.05 & 11.12 & $-$10.86 & -- & 0.95 & -- & 1.94 & 42.1 & (3) \\
GJ~4196 & 1 & M1 & 0.56$\pm$0.05 & 11.06 & $-$67.70 & -- & 0.95 & -- & 2.38 & 42.9 & (3) \\
NLTT~10614 & 1 & M1.5 & 0.54$\pm$0.05 & 11.22 & 5.88 & -- & 0.95 & -- & 2.20 & 36.6 & (3) \\
NLTT~4188 & 1 & M0.5 & 0.59$\pm$0.06 & 11.42 & 8.30 & -- & 0.95 & -- & 2.48 & 41.2 & (3) \\
NLTT~52021 & 1 & M2 & 0.50$\pm$0.05 & 11.56 & $-$27.59 & -- & 1.76 & -- & 2.34 & 29.9 & (3) \\
TYC~1795-941-1 & 1 & M0 & 0.62 & 11.28 & $-$21.33 & -- & 0.95 & -- & 4.99 & 41.0 & (3), (5) \\
TYC~2710-691-1 & 1 & K7.5 & 0.65$\pm$0.07 & 11.91 & $-$10.99 & -- & 1.76 & -- & 4.16 & 39.4 & (3) \\
CCDM~J22441+4030A & 0 & M1.5 & 0.49 & 11.55 & -- & -- & 1.76 & -- & 4.18 & -- & (3), (4), (5) \\
CCDM~J22441+4030B & 0 & M2 & 0.45 & 11.57 & -- & -- & 1.76 & -- & 4.18 & -- & (3), (4), (5) \\
G~243-30 & 0 & M0 & 0.62 & 10.37 & -- & -- & 0.85 & -- & 1.46 & -- & (3), (4), (5) \\
GJ~4120 & 0 & M0 & 0.62 & 10.96 & -- & -- & 1.00 & -- & 1.89 & -- & (3), (4), (5) \\
GJ~4173 & 0 & M0 & 0.62 & 11.21 & -- & -- & 0.95 & -- & 2.48 & -- & (3), (4), (5) \\
GJ~835 & 0 & M0 & 0.62 & 9.83 & -- & -- & 0.72 & -- & 1.27 & -- & (3), (4), (5) \\
NLTT~51676 & 0 & M1 & 0.50 & 10.17 & -- & -- & 0.85 & -- & 1.46 & -- & (3), (4), (5) \\
TYC~3997-1056-1 & 0 & M1 & 0.50 & 11.60 & -- & -- & 1.76 & -- & 4.18 & -- & (3), (4), (5) \\
\hline \hline
\end{longtable}

\end{document}